\documentclass[10pt]{article}

\usepackage{amsmath}
\usepackage{amssymb}
\usepackage{longtable}
\pdfoutput=1

\usepackage{graphicx}
\usepackage{subfigure}
\usepackage{url}

\usepackage{cite}

\usepackage{color} 

\usepackage[labelfont=bf,labelsep=period,justification=raggedright]{caption}

\makeatletter
\renewcommand{\@biblabel}[1]{\quad#1.}
\makeatother

\date{}

\begin{document}

\begin{flushleft}
{\Large
\textbf{Inference of RNA Polymerase II Transcription Dynamics from Chromatin Immunoprecipitation Time Course Data}
}
\\
Ciira wa Maina$^{1,\ast}$,
Antti Honkela$^{2}$,
Filomena Matarese$^{3}$,
Korbinian Grote$^{4}$,
Hendrik G. Stunnenberg $^{3}$, 
George Reid$^{5}$,
Neil D. Lawrence$^{6,\ast}$, 
Magnus Rattray$^{7,\ast}$
\\
\bf{1} Department of Electrical and Electronic Engineering, Dedan Kimathi University of Technology, Nyeri, Kenya
\\
\bf{2} Helsinki Institute for Information Technology HIIT, Department of Computer Science, University of Helsinki, Helsinki, Finland
\\
\bf{3} Nijmegen Centre for Molecular Life Sciences, Radboud University Nijmegen, NL
\\
\bf{4} Genomatix Software GmbH, Muenchen, Germany
\\
\bf{5} Institute for Molecular Biology, Mainz, Germany
\\
\bf{6} Department of Computer Science, University of Sheffield, Sheffield, UK
\\
\bf{7}  Faculty of Life Sciences, University of Manchester, Manchester, UK
\\
$\ast$ E-mail: cwamaina.dekut@gmail.com,magnus.rattray@manchester.ac.uk,\\n.lawrence@sheffield.ac.uk
\end{flushleft}

\section*{Abstract}
Gene transcription mediated by RNA polymerase II (pol-II) is a key step in gene expression. The dynamics of pol-II moving along the transcribed region influence the rate and timing of gene expression.
In this work we present a probabilistic model of transcription dynamics which is fitted to pol-II occupancy time course data measured using ChIP-Seq. The model can be used to estimate transcription speed and to infer the temporal pol-II activity profile at the gene promoter. Model parameters are estimated using either maximum likelihood estimation or via Bayesian inference using Markov chain
Monte Carlo sampling. The Bayesian approach provides confidence intervals for parameter estimates and allows the use of priors that capture domain knowledge, e.g. the expected range of transcription speeds, based on
previous experiments. The model describes the movement of pol-II down the gene body and can be used to identify the time of induction for transcriptionally engaged genes. By clustering the inferred promoter activity time profiles, we are able to determine which genes respond quickly to stimuli and group genes that share activity profiles and may therefore be co-regulated.
We apply our methodology to biological data obtained using ChIP-seq to measure pol-II occupancy genome-wide when MCF-7 human breast cancer cells are treated with estradiol (E2). The transcription speeds we obtain agree with those obtained previously for smaller numbers of genes with the advantage that our approach can be applied genome-wide. We validate the biological significance of the pol-II promoter activity clusters by investigating cluster-specific transcription factor binding patterns and determining canonical pathway enrichment. We find that rapidly induced genes are enriched for both estrogen receptor alpha (ER$\alpha$) and FOXA1 binding in their proximal promoter regions. 

\section*{Author Summary}

Cells express proteins in response to changes in their environment so as to maintain normal function. An initial step in the expression of proteins is transcription which is mediated by RNA polymerase II (pol-II). To understand changes in transcription arising due to stimuli it is useful to model the dynamics of transcription. We present a probabilistic model of pol-II transcription dynamics that can be used to compute RNA transcription speed and infer the temporal pol-II activity at the gene promoter. The inferred promoter activity profile is used to determine genes that are responding in a coordinated manner to stimuli and are therefore potentially co-regulated. Model parameters are inferred using data from 
high-throughput sequencing assays, such as ChIP-Seq and GRO-Seq, and can therefore be applied genome-wide in an unbiased manner. We apply the method to pol-II ChIP-Seq time course data from breast cancer cells stimulated by estradiol in order to uncover the dynamics of early response genes in this system. 

\section*{Introduction}
Transcription mediated by RNA polymerase II (pol-II) is an essential process in the
expression of protein-coding genes in eukaryotes. Transcription is dependent upon a number
of sequential and dynamic events, such as recruitment of pol-II to the transcriptional start site,
activation of pol-II through phosphorylation of its C-terminal domain, elongation of the nascent
transcript through the transcribed region and termination
 \cite{hager2009transcription}.  Each of these steps may be rate-limiting and can therefore affect the level of gene expression. In this manuscript, we describe
a simple probabilistic model of transcription whose parameters can be inferred using time-series data such as pol-II ChIP-Seq data~\cite{Welboren_2009} or nascent transcript measurement by
GRO-Seq that reports markers of
transcriptional activity~\cite{Hah_2011}. This model can be used to identify transcriptionally engaged genes, estimate their
transcription rates and infer transcriptional activity adjacent to the promoter. The transcriptional
dynamics of estrogen responsive genes in a breast cancer cell line were described by fitting this
model to pol-II ChIP-seq time course datasets.

Chromatin immunoprecipitation, in conjunction with massively parallel sequencing (ChIP-seq)
evaluates interactions between proteins and DNA, and, for example, can be used to monitor the
presence of pol-II on DNA. Estimating the amount of pol-II associated with a transcribed gene
provides a measure of transcriptional activity 
 \cite{Welboren_2009}. Sequential measurement of pol-II occupancy on
genes released from transcriptional blockade, for example, in response to stimuli, reveal a wave of
transcription moving through the body of the responding transcript. 

A number of studies have attempted to determine the rate of transcription through modelling the
dynamics of pol-II. Darzacq
 \textit{et al.} fit a mechanistic model of pol-II transcription to nascent RNA data at a single locus and obtained a transcription speed of 4.3 kilobases per minute \cite{Darzacq_2007}.
Wada \textit{et al.} activated transcription of genes greater than 100 kbp in length and estimated the transcription
speeds using a model that measures an intronic RNA signal through taking advantage of co-transcriptional splicing. They obtain an average
transcription rate of
 3.1 kbp min$^{-1}$
 \cite{Wada27102009}. 
Singh and Padget (2009) reversibly inhibit transcription to determine
the transcription rate of 9 genes, all of which were greater than 100 kbp which had an average
transcription rate of
 3.79 kbp min$^{-1}$ \cite{Singh_2009}. The data used in these studies have good temporal
resolution (e.g. samples every 7.5 min in \cite{Wada27102009}) and reliably allow fitting of mathematical models or the direct measurement
of transcription speed, however, only for a limited set of long genes. In contrast, high throughput data
sets such as ChIP-Seq, can be used to uncover transcription dynamics genome-wide but typically
have much lower temporal resolution, motivating the development of alternative modelling approaches 
that report genome-wide transcription rates.

One way around the low temporal resolution of typical high-throughput time course data is to
employ a non-parametric model of the biological signals of interest. In many cases we expect these
signals to vary continuously and smoothly in time, when averaged over a cell population, and a
Gaussian process model provides a convenient non-parametric model in such cases~\cite{gpml06}. Gaussian processes
have recently found applications in a range of biological system models 
~\cite{Gao:latent08, Honkela_2010, Kalaitzis:simple11, Liu2012}. 

Here we present a Gaussian process model of transcription dynamics which can be fitted to 
genome-wide pol-II occupancy data
measured using ChIP-Seq. The model describes the movement of pol-II through the
gene body and combines a flexible model of promoter-proximal pol-II activity
with a reliable estimate of transcription speed. By identifying genes which
fit the model well, we provide a useful method to identify actively transcribed genes. The model 
does not assume a constant transcription speed and can therefore identify variable
rates of transcription, for example due to transcriptional pausing. Model parameters are inferred
using either
maximum likelihood (ML) estimation or via Bayesian inference using Markov chain
Monte Carlo (MCMC) sampling. The Bayesian approach provides confidence intervals 
for parameter estimates and can incoporate priors that
capture domain knowledge, e.g. the expected range of transcription speeds, based on
previous experiments.

We fit our model to a pol-II ChIP-Seq time course dataset from MCF7 breast cancer cells
stimulated with estradiol. The model is used to identify the set of transcriptionally engaged genes and 
estimate their mean transcription rate and transcriptional activity near the
promoter. By clustering promoter activity profiles, potential co-regulated groups of genes are identified, particularly
those that respond rapidly to estrogen signalling. Subsequent characterisation of transcription
factor (TF) binding sites in proximity to the promoters of genes within clusters provides a means
of classifying groups of promoters that are responsive to the binding of specific combinations of
TF’s. Additionally, publically available ChIP-Seq datasets of TF profiles from the same system were
used to identify cluster-specific patterns in TF-binding. The rates of transcription estimated by
our model are consistent with the literature
 \cite{Darzacq_2007,Wada27102009} but with the advantage that our method allows
the computation of transcription speeds genome-wide.

Our methodology has a number of advantages. We do not require data with high temporal
resolution, making it feasible to model transcriptional dynamics genome-wide using ChIP-Seq or
GRO-Seq time course data. We infer transcription rates for all genes in an unbiased manner and
by using Bayesian parameter estimation we are able to associate our transcription rate estimates
with confidence intervals. Our model is non-parametric and therefore does not make very strong
assumptions about the temporal changes in transcriptional activity. Fitting the model genome-wide allows us to identify 
and filter out transcripts where pol-II does not travel down the gene body.
This provides a principled method to identify responsive genes, in particular, early acting estrogen
responsive genes in the specific application considered here. Since our model does not enforce a uniform transcription speed over the entire gene body, we can  take into account phenomena such as pol-II pausing which would result in a non-uniform transcription speed. We also use this model to infer
the promoter activity of transcriptionally engaged genes, to identify co-regulated gene modules
downstream of estrogen signalling.

\section*{Methods}
\begin{figure}[ht!]
\centering
\includegraphics[width=1.0\textwidth]{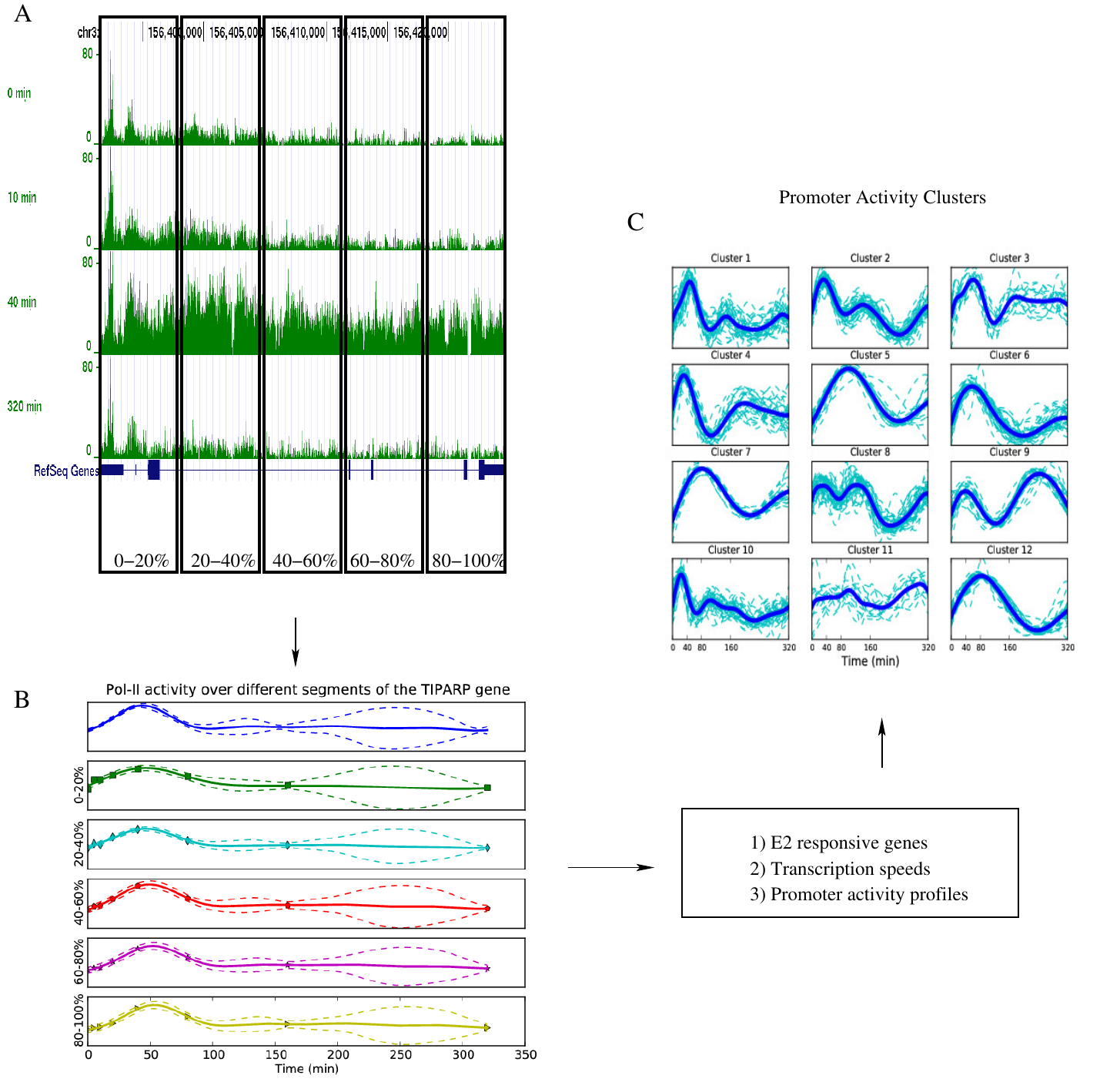}
\caption{Pol-II ChIP-seq data for the TIPARP gene shows a transcription wave moving down the gene. The transcription dynamics model captures this motion and allows us to estimate transcription speeds. In this case the gene is divided into 5 segments and we estimate the speed to be approxiamtely 2 kilobases per minute. Figure A shows the raw ChIP-seq reads at different times between 0 and 320 min. The top panel of Figure B shows the inferred promoter activity profile. The next five panels show the inferred profiles for the five gene segments corresponding to $0-20\%,\ldots,80\%-100\%$ of the gene. By clustering these promoter activity profiles as shown in Figure C, we are able to group genes into clusters that are likely to be co-regulated and in particular we identify the clusters that respond most rapidly to estrogen signalling.
}
\label{fig:schematic}
\end{figure}

Visualizing pol-II ChIP-seq reads mapped to transcriptional units at multiple time points following
the addition of estradiol to MCF7 cells reveals the motion of pol-II through the gene body of
estrogen responsive genes
 (see Figure \ref{fig:schematic}). Computing the average pol-II occupancy over
successive gene segments describes the motion of the transcription wave. Thereafter, fitting a
model capable of smoothly interpolating between observed time points and by determining the time
taken for pol-II to move from one gene segment to the next determines if pol-II is transcriptionally
engaged on a given transcript and the speed at which it is moving through this transcriptional unit.
We use a convolved Gaussian process to model the relationship between the pol-II signal at different regions of the gene and across time. Model parameters are determined using maximum likelihood (ML) or Bayesian
inference via Markov chain Monte Carlo (MCMC) to determine genes of interest and moreover, in the case of MCMC,
determine confidence intervals for our parameter estimates.

\subsection*{Convolved Gaussian Process Model}
A Gaussian process (GP) is a distribution over the space of functions. This distribution is completely specified by a mean function $m(t)$ and a covariance function $k(t,t')$. A function $f(t)$ is said to be drawn from a Gaussian process $\mathcal{GP}(m(t),k(t,t'))$ if $f(t)$ at any finite collection of points has a multivariate Gaussian distribution with mean vector and covariance matrix specified by $m(t)$ and $k(t, t')$, respectively. GPs provide a powerful framework for non-parametric regression~\cite{gpml06}. If a function is assumed to be drawn from a GP with known mean and covariance function, we can infer the function value and associated uncertainty at unobserved locations given noise-corrupted observations. GPs have recently been applied in modelling biological systems,  e.g. modelling protein concentrations as latent variables in differential equation models of transcriptional regulation~\cite{Gao:latent08, Honkela_2010} and modelling spatial gene expression~\cite{Liu2012}. 

Here we introduce a novel application of GPs to modelling the spatio-temporal dynamics of pol-II occupancy during transcription. Convolved GPs allow the modelling of correlations between multiple coupled data sources. In our case these data sources are the pol-II occupancy over time collected at different locations along the transcribed region of a gene. Modelling the data as a convolved process borrows information from these different data sources in estimating the model parameters and inferring the underlying signal in the data.  Also, we find that convolved GPs are necessary to account for changes in the shapes of signals observed at different regions of the gene. In linear systems theory, the output $y(t)$ of a linear time-invariant system whose impulse response is $h(t)$ is given by the convolution of the input $x(t)$ and $h(t)$, that is $y(t)=\int_{-\infty}^{\infty}h(\tau)x(t-\tau)d\tau$. If different sets of observations are believed to be related, they can be modeled as the outputs of different linear systems in response to a single input. If this input is modeled as a GP, then it will form a joint GP together with all the outputs and data from one output stream will be useful in inferring the rest \cite{Higdon:convolutions02,Higdon01,Higdon:ocean98,Boyle05,Alvarez08,Alvarez:vector12,Alvarez:computationally11,verHoef:convolution98,Calder:convolution07}. In our case, incorporating the data from multiple spatially separated regions of the genes allows us to infer an underlying function that links all these regions. This proves useful as a summary of the transcription dynamics of the gene and we show that it provides useful insights into potential coregulation.

\subsubsection*{Model Description}

In order to capture the movement of the transcription wave through transcriptional units, we
divide each gene into $I$ segments and compute time series of pol-II occupancy for each of
the segments. Due to the low temporal resolution characteristic of high-throughput datasets,
the time series between measurements must be inferred.
 To this end, we model the pol-II occupancy $y_i(t)$ in each segment $i\in\{1,\ldots,I\}$ as the convolution of a latent process $f(t)$ which is shared by all segments and a (possibly delayed) smoothing kernel $k_i(\tau-D_i)$ corrupted by an independent white Gaussian noise process $\epsilon_i(t)$ with zero mean and variance $\sigma_i^2$ \cite{Boyle05,Alvarez08}. That is 
\begin{equation}
\label{eq:obs}
y_i(t)=\alpha_i\int_{-\infty}^{\infty}f(t-\tau)k_i(\tau-D_i)d\tau+\epsilon_i(t),
\end{equation}
where $\alpha_i$ is a scale factor and $D_i$ is the delay of each segment.
The latent process $f(t)$ is modeled as a random function drawn from a GP with zero mean and a squared exponential covariance function (defined in Equation~(\ref{eq:rbf}) below). The smoothing kernel is assumed to be Gaussian, that is
\begin{equation}
k_i(\tau)=\frac{1}{\sqrt{2\pi}\ell_i}\exp\Bigg(-\frac{\tau^2}{2\ell_i^2}\Bigg).
\end{equation}
The estimated delay $D_i$ of each smoothing kernel models the amount of time it takes the `transcription wave' to reach the corresponding gene segment. This is used to estimate the transcription speed. Biologically the latent function can be thought of as modeling activity at the promoter while the smoothing kernel accounts for `diffusion' of the transcription wave. This diffusion phenomenon is observed when time series of pol-II occupancy over different sections of a gene are plotted, with the transcription wave seen to spread out (see Figure \ref{fig:twave_lg}).  This phenomenon may be due to an initially synchronized cell population becoming less synchronized over time, resulting in broadening of the pol-II occupancy distribution over time. The parameter $\ell_i$ captures the amount of `spread' observed at the $i$th segment. It also serves as a measure of the loss of synchrony between the cells of the population when the transcription wave is observed at the $i$th segment.

Using equation (\ref{eq:obs}), we can compute the covariance between the pol-II occupancy at various segments of the gene. We have 
\begin{equation}
\label{eq:obsCov}
\mathsf{cov}[y_i(t),y_j(t')]=\alpha_i\alpha_j\int_{-\infty}^{\infty}\int_{-\infty}^{\infty}k_{f}(t-\tau,t'-\tau')k_i(\tau-D_i)k_j(\tau'-D_j)d\tau d\tau' + \sigma_i^2\delta_{ij}\delta_{tt'}
\end{equation}
where
\begin{equation}
\label{eq:rbf}
k_{f}(t,t')=\sigma_f^2\exp\Bigg(-\frac{(t-t')^2}{2\ell_f^2}\Bigg).
\end{equation}
Equation (\ref{eq:obsCov}) can be evaluated in closed form using the fact that the product of two Gaussians yields an un-normalized Gaussian \cite{gpml06}. Exploiting this fact we get
\begin{equation}
\label{eq:obsCov2}
\mathsf{cov}[y_i(t),y_j(t')]=\alpha_i\alpha_j\frac{\sigma_f^2\ell_f}{\sqrt{\ell_f^2+\ell_i^2+\ell_j^2}}\exp\Bigg(-\frac{(t'-t+D_i-D_j)^2}{2(\ell_f^2+\ell_i^2+\ell_j^2)}\Bigg) + \sigma_i^2\delta_{ij}\delta_{tt'}.
\end{equation}
Similarly,
\begin{equation}
\label{eq:obsCov3}
\mathsf{cov}[f(t),y_i(t')]=\alpha_i\frac{\sigma_f^2\ell_f}{\sqrt{\ell_f^2+\ell_i^2}}\exp\Bigg(-\frac{(t'-t-D_i)^2}{2(\ell_f^2+\ell_i^2)}\Bigg).
\end{equation}

\subsubsection*{Parameter Estimation and Inference}
Let $\mathbf{y}_i=[y_{i1},\ldots,y_{iN}]^\top$ be a vector of observations of pol-II occupancy over the i$th$ gene segment and let $\mathbf{Y}=[\mathbf{y}_1^\top,\ldots,\mathbf{y}_I^\top]^\top$ be a vector formed by concatenating all the observations for a single gene. $N$ is the number of observation time points and $I$ is the number of gene segments so for a single gene $\mathbf{Y}$ is a vector of length $NI$. We have
\begin{equation}
p(\mathbf{f},\mathbf{Y}|\Theta)=\mathcal{N}([\mathbf{f},\mathbf{Y}]; \mathbf{0},\mathbf{K}),
\end{equation} 
where
\begin{equation}
\label{eqn:covmtx}
\mathbf{K} = \left[
\begin{array}{cccc}
\mathbf{K}_{\mathbf{f,f}} & \mathbf{K}_{\mathbf{f,y_1}} & \ldots&\mathbf{K}_{\mathbf{f,y_I}} \\
\mathbf{K}_{\mathbf{y_1,f}} & \mathbf{K}_{\mathbf{y_1,y_1}} & \ldots &\mathbf{K}_{\mathbf{y_1,y_I}} \\
\vdots & \vdots & \ddots&\vdots\\
\mathbf{K}_{\mathbf{y_I,f}}&\ldots&\ldots&\mathbf{K}_{\mathbf{y_I,y_I}}
\end{array} \right]
\end{equation}
and $\Theta=\{\sigma_f,\ell_f,\{\alpha_i,D_i,\ell_i,\sigma_i\}_{i=1}^I\}$ are the parameters of our model which will be fitted on a gene by gene basis. The elements of $\mathbf{K}$ are computed using equations (\ref{eq:rbf}), (\ref{eq:obsCov2}), and (\ref{eq:obsCov3}). By marginalizing over the latent function $\mathbf{f}$, we obtain the marginal likelihood $p(\mathbf{Y}|\Theta)$. Maximum likelihood estimates of the parameters $\Theta$ are readily obtained by maximizing the log marginal likelihood using gradient-based optimisation.

For a fully Bayesian approach, we take advantage of the fact that the parameters are positive and bounded. We transform the parameters using a logit transform and work with unconstrained variables. We place a Gaussian prior over the parameters in the transformed domain and draw samples from the posterior using the Hamiltonian Monte Carlo (HMC) algorithm \cite{Neal_HMC} (A more detailed description of the priors is included in the supplementary material). 

Code to implement the method is freely available as a Python package, PyPol-II, which can be downloaded from \url{https://github.com/ciiram/PyPol_II}.

\subsubsection*{Estimation of Average Transcription Speed}

When fitting the model, we fix $D_1=0$ to ensure identifiability.
The average transcription speed is computed by assuming that the value of $D_i$ is an indicator of how long it takes the `transcription wave' to reach the corresponding gene segment. That is, $D_2$ is the amount of time it takes to transcribe 20\% of the gene, $D_3$ 40\% etc.  
To obtain confidence intervals on the delay estimates, MCMC was performed to get samples of the parameters.

To compute the average transcription speed we plot the position along the gene in base pairs (bp) versus the delay in minutes and compute a linear regression through the origin. The slope of the regression line gives us the transcriptional speed. Each sample of the parameters provides a set of delay estimates from which we obtain a speed estimate.

\subsection*{Alternative Methods for Time Delay Inference}
A key component of our method involves the estimation of delay between time series observed at different segments of the gene. The study of time delay between related time series has received attention from a number of researchers for a long time \cite{Knapp_1976}. The application areas range from signal processing to astronomy \cite{Haarma_99}. The classic approach to time delay estimation involves computing the cross-correlation between the related time series and determining the value of delay for which this function is maximised.
Consider two signals $y_1(t)$ and $y_2(t)$ given by
\begin{eqnarray}
\label{eq:delay_df}
y_1(t)&=&f(t)+n_1(t)\nonumber\\
y_2(t)&=&f(t-D)+n_2(t)
\end{eqnarray}
where $n_1(t)$ and $n_2(t)$ are uncorrelated noise processes. The cross-correlation function is given by $R_{y_1,y_2}(\tau)=\mathbf{E}[y_1(t)y_2(t-\tau)]$ where $\mathbf{E}$ denotes the expectation operator. The value of $\tau$ that maximises $R_{y_1,y_2}(\tau)$ yields an estimate of the delay $D$. When the signals are sampled at $N$ equally spaced time points $t_0,\ldots,t_{N-1}$ with spacing $T$ between samples, the discrete time equivalent of $R_{y_1,y_2}(\tau)$ is readily estimated. Let $y_1[n]=y_1(nT)$, the discrete cross-correlation is estimated as 
\begin{displaymath}
\hat{R}_{y_1,y_2}(kT)=\frac{1}{N}\sum_{n=0}^{N-1-k}y_1[n]y_2[n+k].
\end{displaymath}
The delay is estimated by finding the value of $k$ for which $\hat{R}_{y_1,y_2}(kT)$ is maximised. The corresponding delay estimate is $kT$. 
 However, this approach doesn't work well when the time series are unevenly sampled as is the case in several astronomical and biological studies. A number of techniques have been developed to handle unevenly sampled time series including the discrete correlation function (DCF) \cite{DCF_88}, and the more recent kernel based approaches \cite{Tello_2006,Harva_2008}. The DCF is computed as follows, for all $i,j\in\{0,\ldots,N-1\}$ the time differences $\Delta_{ij}=|t_i-t_j|$ are binned into discrete bins of size $\Delta\tau$. The DCF at $\tau$ is given by \cite{DCF_88,Tello_2006}
\begin{equation}
DCF(\tau)=\frac{1}{|S(\tau)|}\sum_{(i,j) \in S(\tau)}\frac{(y_1[i]-\bar{y_1})(y_2[j]-\bar{y_2})}{\sqrt{(\sigma_{y_1}^2-\sigma_{y_{1i}}^2)(\sigma_{y_2}^2-\sigma_{y_{2j}}^2)}},
\label{eq:dcf}
\end{equation}
where
\begin{equation}
  \label{eq:dcf_clarification}
  S(\tau) = \{(i,j) | \Delta_{ij}\in[\tau-\Delta\tau,\tau+\Delta\tau]\},
\end{equation}
and $\sigma_{y_1}^2$ and $\sigma_{y_2}^2$ are the variances of the observation streams while $\sigma_{y_{1i}}^2$ and $\sigma_{y_{2j}}^2$ are observation error variances.

In the kernel based approach of \cite{Tello_2006}, the underlying function $f(t)$ of equation (\ref{eq:delay_df}) is modelled as the sum of a fixed number of kernels centered at the observation times. That is 
\begin{equation}
f(t)=\sum_{i=0}^{N-1}\alpha_iK(c_i,t)
\end{equation}
where 
\begin{equation}
K(c_i,t)=\exp\Bigg(-\frac{(t-c_i)^2}{\sigma_i^2}\Bigg).
\end{equation}
The value of $D$ that minimises the estimation error is the delay estimate. Our implementation follows that presented in \cite{Tello_2006} where we assumed a fixed kernel width. This kernel width is determined by leave one out cross-validation.



\subsection*{Benchmark Data}

We used synthetic data and previously published experimental data to assess our novel method's performance. To generate the synthetic data, the underlying function $f(t)$ of equation (\ref{eq:delay_df}) was given as a sum of Gaussian kernels. That is 
\begin{displaymath}
f(t)=\sum_{i=1}^N\beta_i\exp\Bigg(-\frac{(t-c_i)^2}{\sigma_i^2}\Bigg).
\end{displaymath} 
N was fixed at 20 and the observation interval $t\in[0,10]$. $\beta_i$, $\sigma_i$ and $c_i$ were generated at random with $\beta_i\in[0,1]$, $\sigma_i\in(0.5,1.5]$ and $c_i\in[2.5,5]$. A random delay $D\in[1,2.5]$ was used to generate the observations which were corrupted by additive Gaussian noise with $\sigma_n=0.001$. To determine the effect of number of observations on the quality of inference we compute the Median Normalised Square Error (MNSE) of the estimated delay $\frac{\Vert D-\hat{D}\Vert_2^2}{\Vert D\Vert_2^2} $ as a function of the number of observations for 50 random realisations of the the signals. We also investigated the effect of distorting the shape of the observed signals by introducing convolution. In real signals the restriction that the shape remains unchanged sometimes leads to poor results. The parameters of the smoothing kernel in equation (\ref{eq:obs}) were generated at random with $\alpha_i\in[0,1]$ and $\ell_i\in(0.625,2.5]$.

To assess performance of our method on a well characterised real-world dataset we obtained a dataset from Singh and Padgett \cite{Singh_2009} where the delay in appearance of pre-mRNA signal at exon-intron junctions was used to compute estimates of transcription speed for 9 genes. To generate the data, transcription was reversibly inhibited \textit{in vivo} using 5,6-dichlorobenzimidazole 1-beta-D-ribofuranoside (DRB) and the pre-mRNA measured after the inhibitor was removed. As verified by the authors, the kinetics of pol-II and pre-mRNA are similar hence we expect good performance on this dataset to indicate applicability of our method to pol-II ChIP-seq data.

\subsection*{pol-II ChIP-Seq Data}

To demonstrate an application to pol-II ChIP-Seq data, we apply our model to investigate the transcriptional response to Estrogen Receptor signalling. ChIP-seq was used to measure pol-II occupancy genome-wide when MCF-7 breast cancer cells are treated with estradiol (E2). Cells were put in estradiol free media for three days. This is defined media devoid of phenol red (which is estrogenic) containing 2\% charcoal stripped foetal calf serum. The charcoal absorbs estradiol but not other essential serum components, such as growth factors. This results in basal levels of transcription from E2 dependent genes. The cells are then incubated with E2 containing media, which results in the stimulation of estrogen responsive genes. The measurements were taken at logarithmically spaced time points 0, 5, 10, 20, ..., 320 minutes after E2 stimulation. 

Raw reads were mapped onto the human genome reference sequence (NCBI\_build37) using the Genomatix Mining Station (software version 3.2.1). The mapping software on the Mining Station is an index based mapper that uses a shortest unique subword index generated from the reference sequence to identify possible read positions. A subsequent alignment step is then used to get the highest-scoring match(es) according to the parameters used.  We used a minimum alignment quality threshold of 92\% for mapping and trimmed 2 basepairs from the ends of the reads to account for deterioration in read quality at the 3' end. The software generates separate output files for uniquely mapped reads and reads that have multiple matches with equal score. We only used the uniquely mapped reads. On average about 66\% of all reads could be mapped uniquely. The data are available from the NCBI Gene Expression Omnibus under accession number GSE44800.

 Time series of pol-II occupancy over various segments of genes were computed in reads per million (RPM) \cite{Pepke_2009} using BEDtools \cite{Quinlan_2010,dale2011pyb}. The genes were divided into 200bp bins and the RPM computed for each bin. The occupancy in a particular gene segment was the mean RPM of the bins in that segment. Here, the gene is divided into five segments each representing 20\% of the gene.

\section*{Results}
\subsection*{Assessment on Benchmark Data}
We first applied our methodology to synthetic data in order to compare its performance to other methods. We investigated the performance of five methods, namely cross-correlation (Corr), DCF, the kernel approach of \cite{Tello_2006} (Kern), a GP approach with no convolution (GP-NoConv), and the convolved GP approach developed in this paper (GP-Conv).  Tables \ref{tab:toy_data1} and \ref{tab:toy_data2} show the MNSE for the different delay estimation methods as a function of the number of observations for synthetic data without convolution and with convolution respectively. Note that the kernel and DCF methods require an estimate of the noise variance and in this simulation study we provide the algorithms with the true value, but that would not be known in practice. We see that when no convolution is introduced, the kernel method performs well but is outperfomed by both GP methods. When convolution is introduced the kernel method appears to break down and as expected the GP-Conv outperforms the other techniques. 

\begin{figure}[ht!]
\centering
\includegraphics[width=0.75\textwidth]{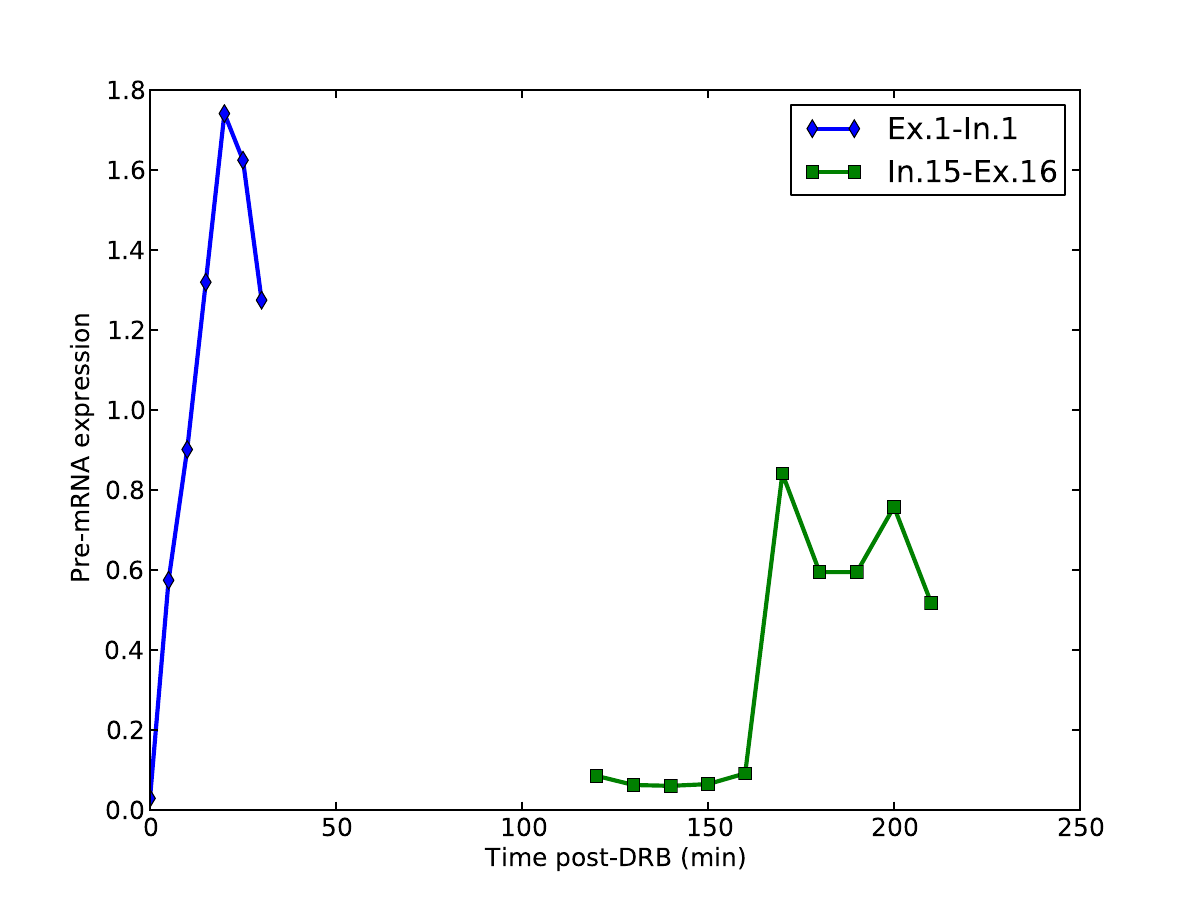}
\caption{Pre-mRNA espression at exon-intron junctions for the \textit{SLC9A9} gene. 
}
\label{fig:pre-mRNA_1}
\end{figure}

\begin{figure}[ht]
\centering
\subfigure[]{
\includegraphics[width=0.48\textwidth]{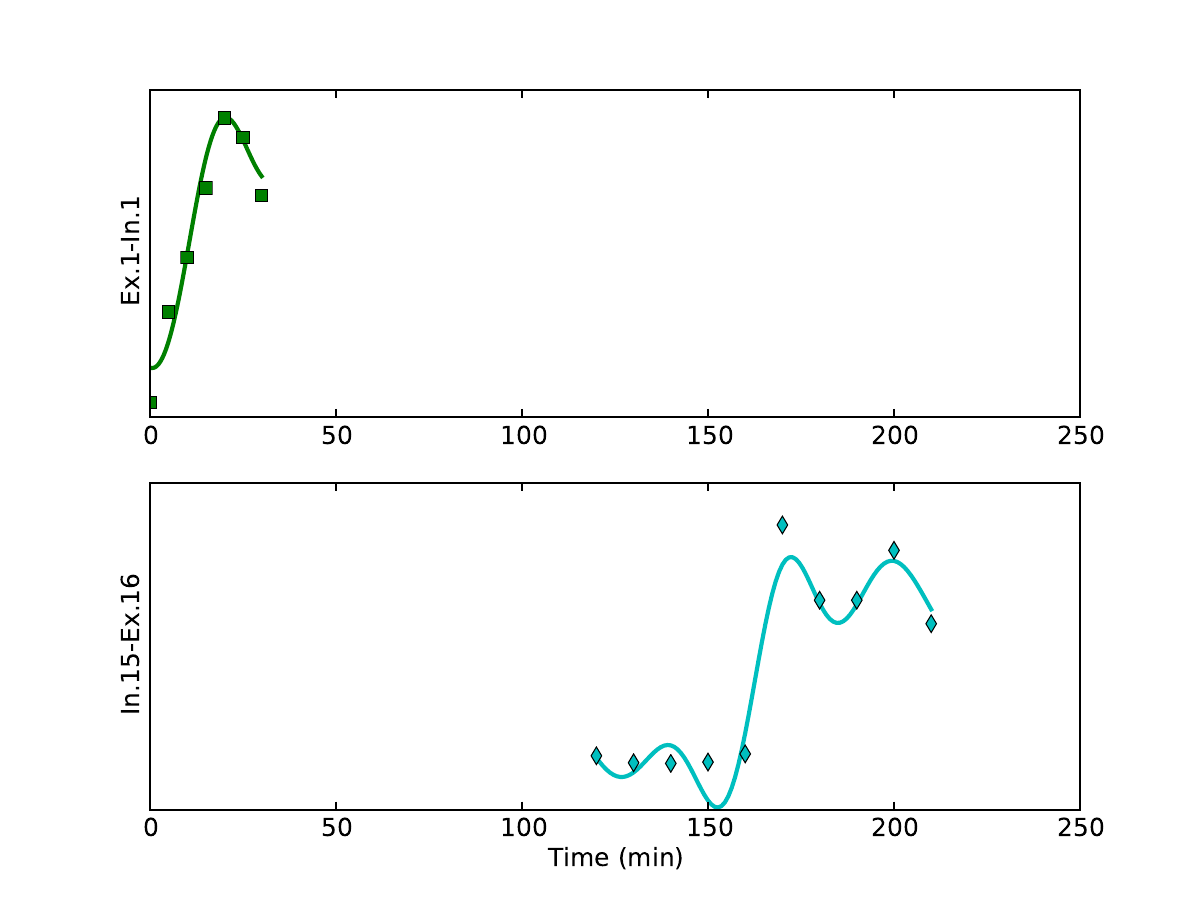}
\label{fig:SLC9A9_1}
}
\subfigure[]{
\includegraphics[width=0.48\textwidth]{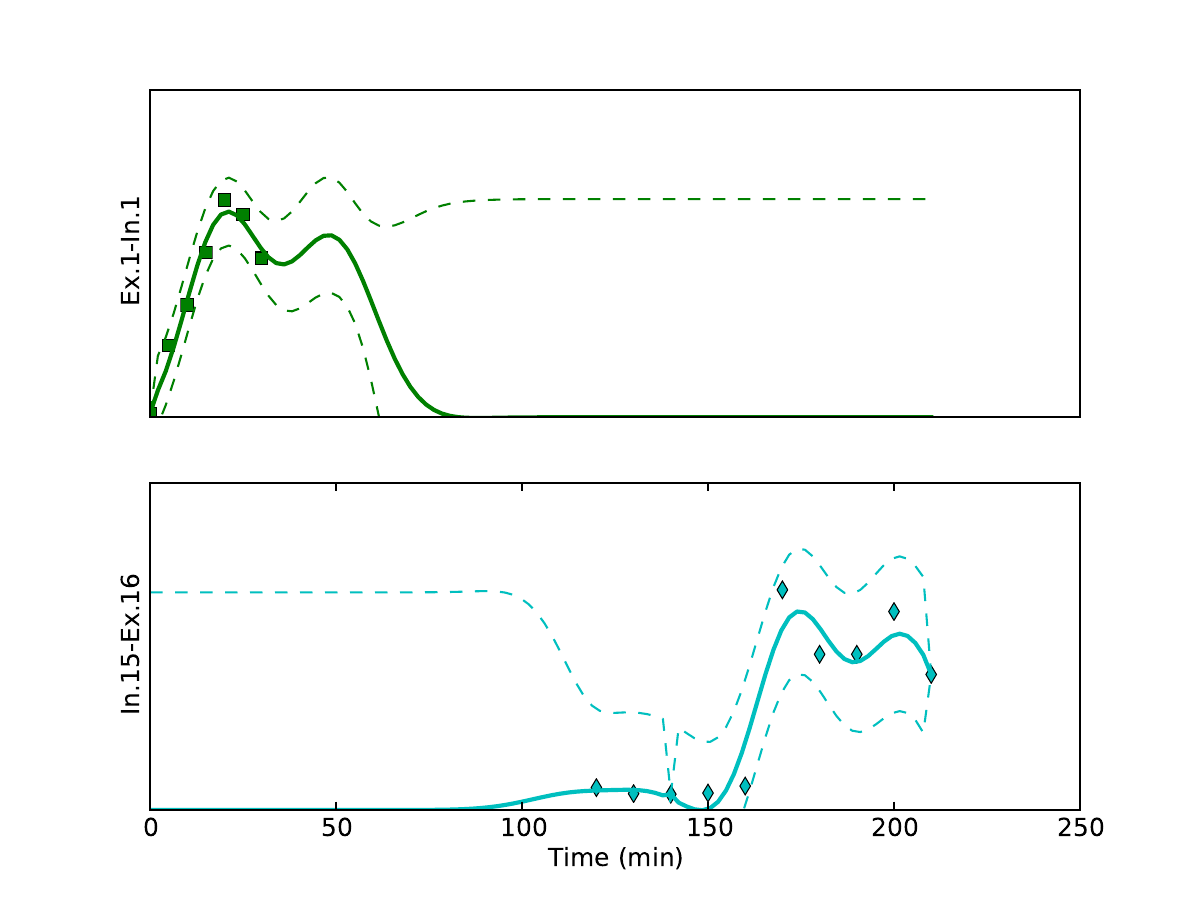}
\label{fig:SLC9A9_2}
}
\subfigure[]{
\includegraphics[width=0.48\textwidth]{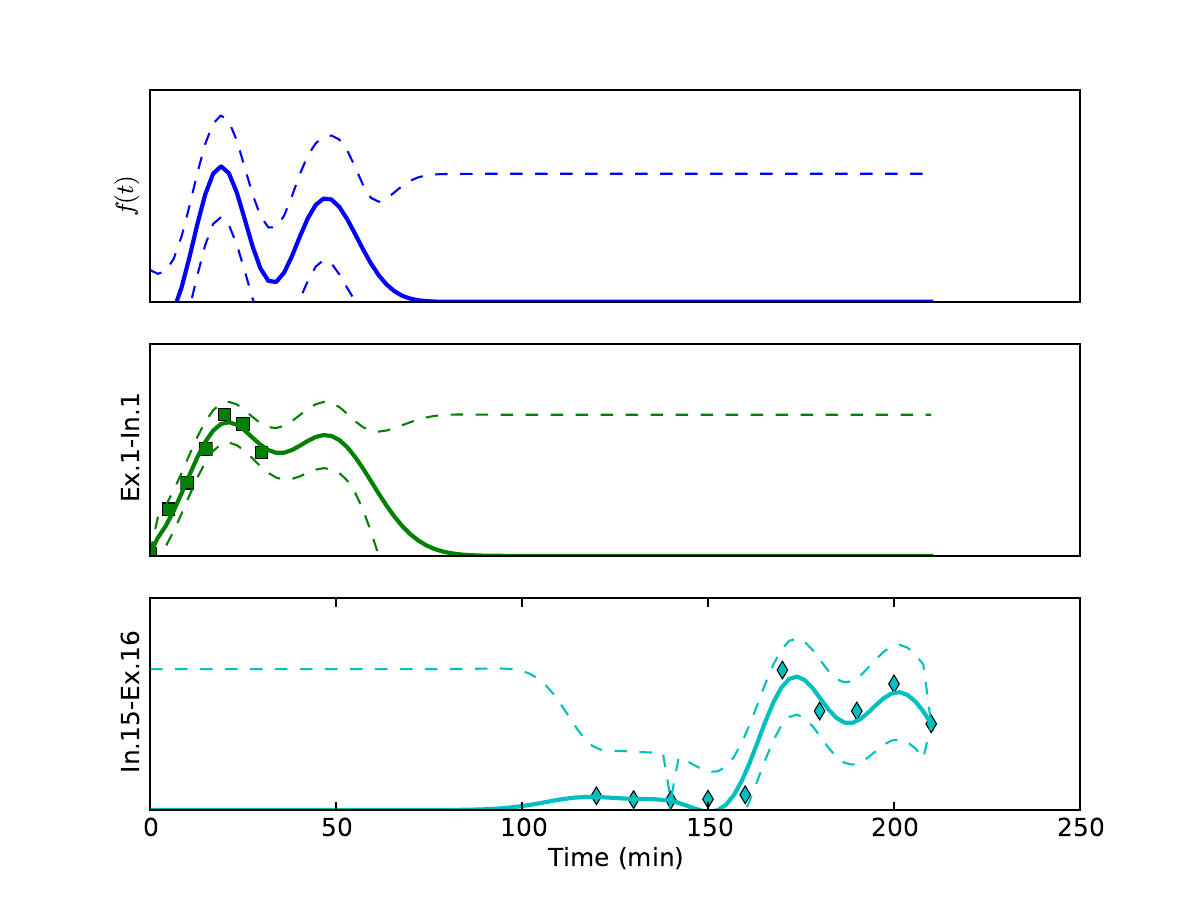}
\label{fig:SLC9A9_3}
}
\caption[Optional caption for list of figures]{Fits for the \textit{SLC9A9} gene using the kernel method (a) and the two GP methods: GP\_NoConv (b) and GP\_Conv (c). In the GP case we show the 95\% confidence interval using dashed lines. In regions with no observations, the uncertainty is large. 
 }
\label{fig:pre-mRNA_2}
\end{figure}

We next applied the model to pre-mRNA data from Singh and Padgett \cite{Singh_2009} where the delay in appearance of pre-mRNA signal at exon-intron junctions was used to compute estimates of transcription speed for 9 genes.
Figure \ref{fig:pre-mRNA_1} shows the pre-mRNA signal for the \textit{SLC9A9} gene (the same data shown in Figure 4d of  \cite{Singh_2009}). The delays read from these plots were used in  \cite{Singh_2009} to determine transcription speeds. Figure \ref{fig:pre-mRNA_2} shows the fit obtained using the kernel method, GP-NoConv and GP-Conv respectively. Table \ref{tab:singh_data} shows the delays read off the plots as well as values obtained using the five delay estimation algorithms for different regions of the nine genes presented in \cite{Singh_2009}. In each row the delay estimate with the lowest normalised square error is highlighted. Table \ref{tab:singh_data2} shows the MNSE for the five delay estimation algorithms for all the genes. We see that the convolved GP method developed in this paper outperforms the other techniques. This method has the added advantage of inferring a latent function which links all the observations and which can be used for downstream analysis. Also, when analysis is genome-wide, reading delays off individual plots is not feasible and furthermore when the sampling intervals are irregularly spaced assigning delays manually would be error prone. These results serve to justify the use of the convolved GP method introduced in this paper.

\begin{center}
\begin{table}[ht]
\begin{center}
\begin{tabular}{|c|c|c|c|c|c|} 
\hline 
Number of  &\multicolumn{5}{|c|}{MNSE}\\ \cline{2-6}  
Observations&Corr&DCF&Kern \cite{Tello_2006}&GP-NoConv&GP-Conv\\
\hline
6 & 36e-3 & 30e-3 & 4e-3 & 1.6e-3 & 2.2e-3 \\
\hline
8 & 44e-3 & 48e-3 & 1.0e-3 & 0.16e-3 & 0.17e-3 \\
\hline
10 & 11e-3 & 13e-3 & 1.2e-3 & 0.0076e-3 & 0.012e-3 \\
\hline
12 & 19e-3 & 18e-3 & 1.2e-3 & 0.0018e-3 & 0.0014e-3 \\
\hline
\end{tabular} 
\end{center}
\caption{MNSE as a function of the number of observations with no convolution.}
\label{tab:toy_data1}
\end{table} 
\end{center}

\begin{center}
\begin{table}
\begin{center}
\begin{tabular}{|c|c|c|c|c|c|} 
\hline 
Number of  &\multicolumn{5}{|c|}{MNSE}\\ \cline{2-6}  
Observations&Corr&DCF&Kern \cite{Tello_2006}&GP-NoConv&GP-Conv\\
\hline
6 & 32e-3 & 37e-3 & 17000e-3 & 0.16e-3 & 0.053e-3\\
\hline
8 & 57e-3 & 61e-3 & 16000e-3 & 0.098e-3 & 0.0057e-3\\
\hline
10 & 11e-3 & 15e-3 & 17000e-3 & 0.018e-3 & 0.0021e-3\\
\hline
12 & 22e-3 & 31e-3 & 23000e-3 & 0.028e-3 & 0.011e-3\\
\hline
\end{tabular} 
\end{center}
\caption{MNSE as a function of the number of observations with convolution.}
\label{tab:toy_data2}
\end{table} 
\end{center}


\begin{center}
\begin{table}[ht!]
\begin{center}
\begin{tabular}{|c|c|c|c|c|c|c|c|c|} 
\hline 
Gene &Region &Length&Delay&Corr&DCF&Kern&GP&GP\\  
&&(kb)&(min)\cite{Singh_2009}&&&\cite{Tello_2006} &NoConv&Conv\\
\hline
Utrophin & Ex1-Ex2 & 111 & 30&15.0 &10.8& 3.1&46.9&\textbf{17.4}\\
\hline
Utrophin & Ex2-Ex50 & 174 & 40 &-&49.2& 125.5&49.5&\textbf{46.8}\\
\hline
Utrophin & Ex50-Ex51 & 101 &25 &-&10.8&67.3&\textbf{34.1}&13.8\\
\hline
Utrophin & Ex51-Ex74 & 173 & 40 &-&238.3&214.3&9.9&\textbf{68.5}\\
\hline
Utrophin & Ex1-Ex74 & 561&140  &-&135.6& 128.6&\textbf{140.3}&146.4\\
\hline
ITPR1&Ex1-Ex5&133&40&45.0&45.5&\textbf{41.3}&49.2&43.2\\
\hline
ITPR1&Ex5-Ex40&105&25&\textbf{25.0}&24.8&23.0&17.4&24.0\\
\hline
ITPR1&Ex1-Ex40&238&65&70.0&69.8&96.4&\textbf{66.6}&67.2\\
\hline
EFNA5&Ex1-Ex2&243&70&65.0&65.4&146.9&69.8&\textbf{69.9}\\
\hline
BCL2&Ex2-Ex3&189&50&5.0&\textbf{54.9}&81.3&65.0&55.0\\
\hline
OPA1&Ex1-Ex29&104&25&20.0&\textbf{25.0}&14.9&27.0&26.8\\
\hline
IFT80&Ex1-Ex20&142&35&40&74.6&\textbf{35.2}&41.6&41.6\\
\hline
CTNNBL1&Ex1-Ex16&178&45&\textbf{45.0}&45.4&39.1&47.2&47.1\\
\hline
KIFAP3&Ex1-Ex20&153&45&\textbf{45.0}&45.4&39.1&46.7&46.7\\
\hline
SLC9A9&Ex1-Ex16&583&160&-&150.2&152.0&\textbf{153.6}&153.5\\
\hline
\end{tabular} 
\end{center}
\caption{Transcription time estimates for different delay estimation algorithms using the pre-mRNA data from \cite{Singh_2009}. When sampling times are uneven, cross-correlation results are omitted. In each row the delay estimate with the lowest normalised square error is highlighted.} 
\label{tab:singh_data}
\end{table} 
\end{center}

\begin{center}
\begin{table}
\begin{center}
\begin{tabular}{|c|c|c|c|c|c|} 
\hline 
&Corr&DCF&Kern \cite{Tello_2006}&GP-NoConv&GP-Conv\\
\hline
MNSE&0.115 &1.787 & 1.974&0.090 &\textbf{0.065}\\
\hline
\end{tabular} 
\end{center}
\caption{MNSE for the 5 delay estimation algorithms for all the genes using pre-mRNA data.}
\label{tab:singh_data2}
\end{table} 
\end{center}

\subsection*{Application to Estrogen Response ChIP-Seq Data}

We applied our method to a ChIP-Seq time-course dataset measuring pol-II occupancy genome-wide when MCF-7 cells are treated with estradiol (E2).  For our initial experiment, we considered 3,064 genes which exhibit significant increase
 of pol-II occupancy between 0 and 40 minutes after E2 treatment. These genes were determined by counting the number of pol-II tags on the annotated genes in the RefSeq hg19 assembly at 0 and 40 minutes after E2 treatment and computing the $\mathrm{log}_2$ ratio of these counts. We keep those genes where this quantity is greater than one standard deviation above the mean.  For these 3,064 genes, we filtered out genes less than 1000bp in length and computed model fits using the ChIP-seq time series data for the remaining 2623 genes.
 The estimation of the parameters $\{\sigma_f,\ell_f,\{\alpha_i,D_i,\ell_i,\sigma_i\}_{i=1}^5\}$ for a given gene was performed using maximum likelihood with $D_1$ fixed at zero, $\sigma_f=1$ and the values $\sigma_i$ constrained to be equal. Intuitively, one would expect the values of delay $\{D_i\}_{i=1}^5$ to be non-decreasing. We therefore keep only those genes where this natural ordering is preserved for further analysis. We also discard genes with $\hat{\ell}_f\leq 10$ and $\hat{\ell}_f\geq 200$ 
 since these are generally seen to be poor fits. Small values of $\hat{\ell}_f$ arise when the data is best modelled as a noise process while large values model constant profiles which are not interesting in our analysis. This left us with 383 genes which we consider a conservative set of genes where there is evidence of engaged transcription and where the model parameters can be confidently estimated. To rank these genes we compared the log marginal likelihood of the model fit to that obtained if we assume independence between the segments, which is equivalent to setting the off-diagonal blocks in equation (\ref{eqn:covmtx}) to the zero matrix. 

\begin{figure}[!ht]
\centering
\subfigure[]{
\includegraphics[width=0.38\textwidth]{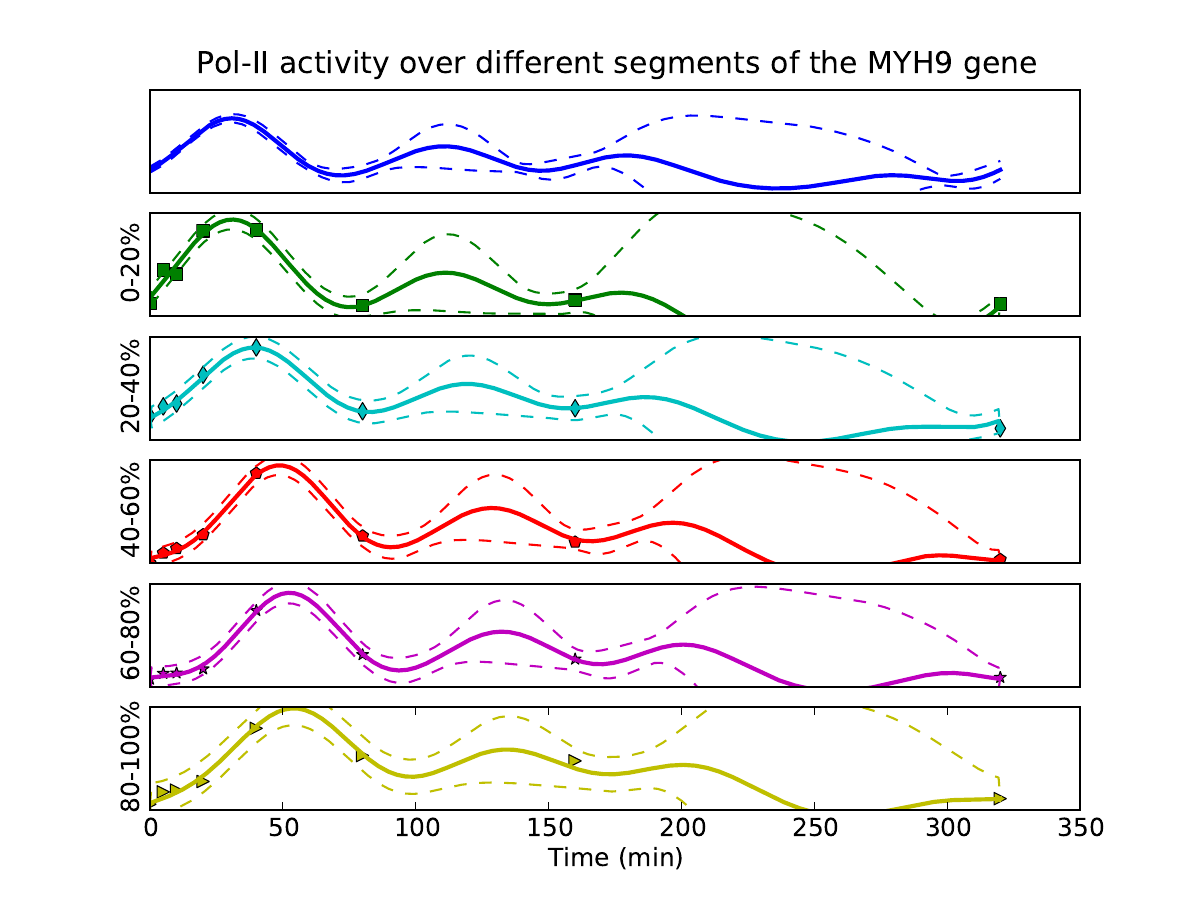}
\label{fig:myh9_2}
}
\subfigure[]{
\includegraphics[width=0.38\textwidth]{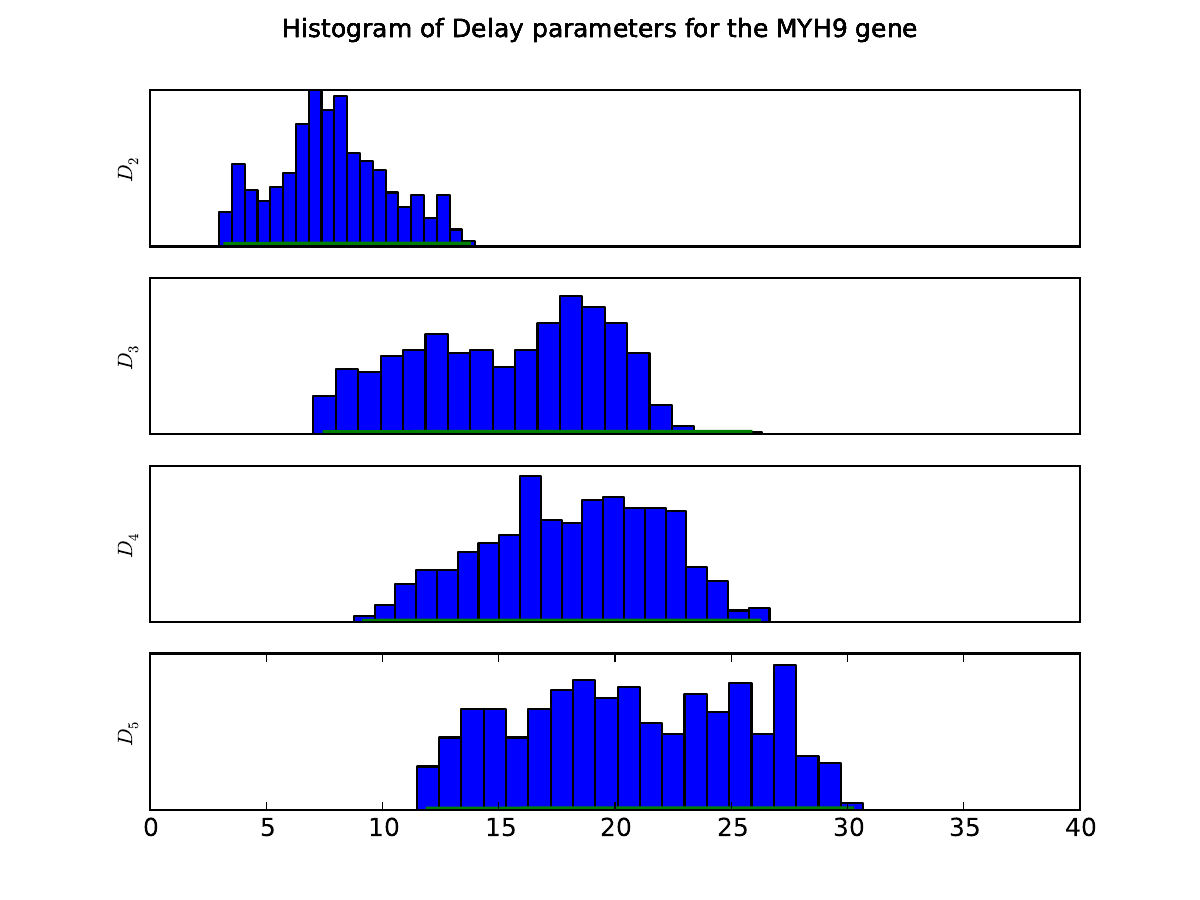}
\label{fig:akap1_2}
}
\subfigure[]{
\includegraphics[width=0.38\textwidth]{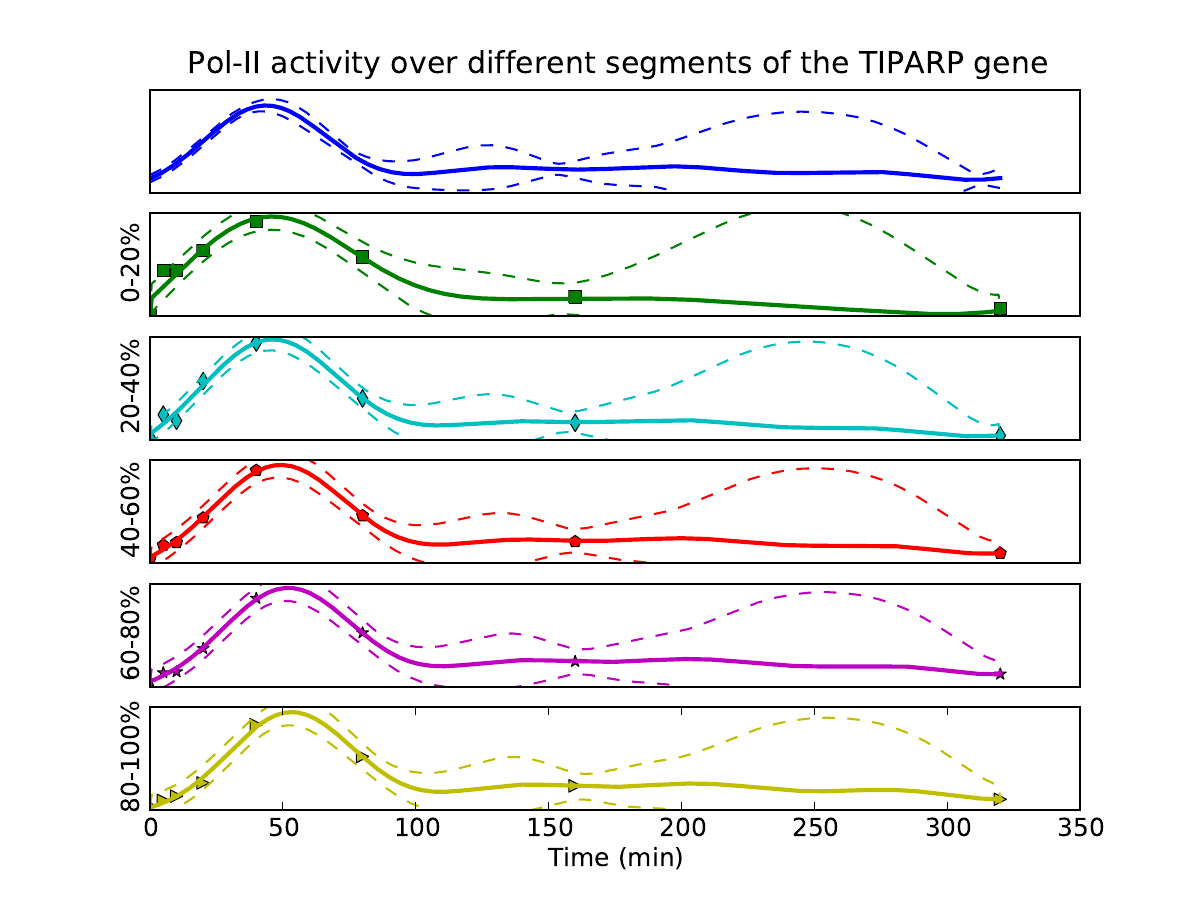}
\label{fig:actn1}
}
\subfigure[]{
\includegraphics[width=0.38\textwidth]{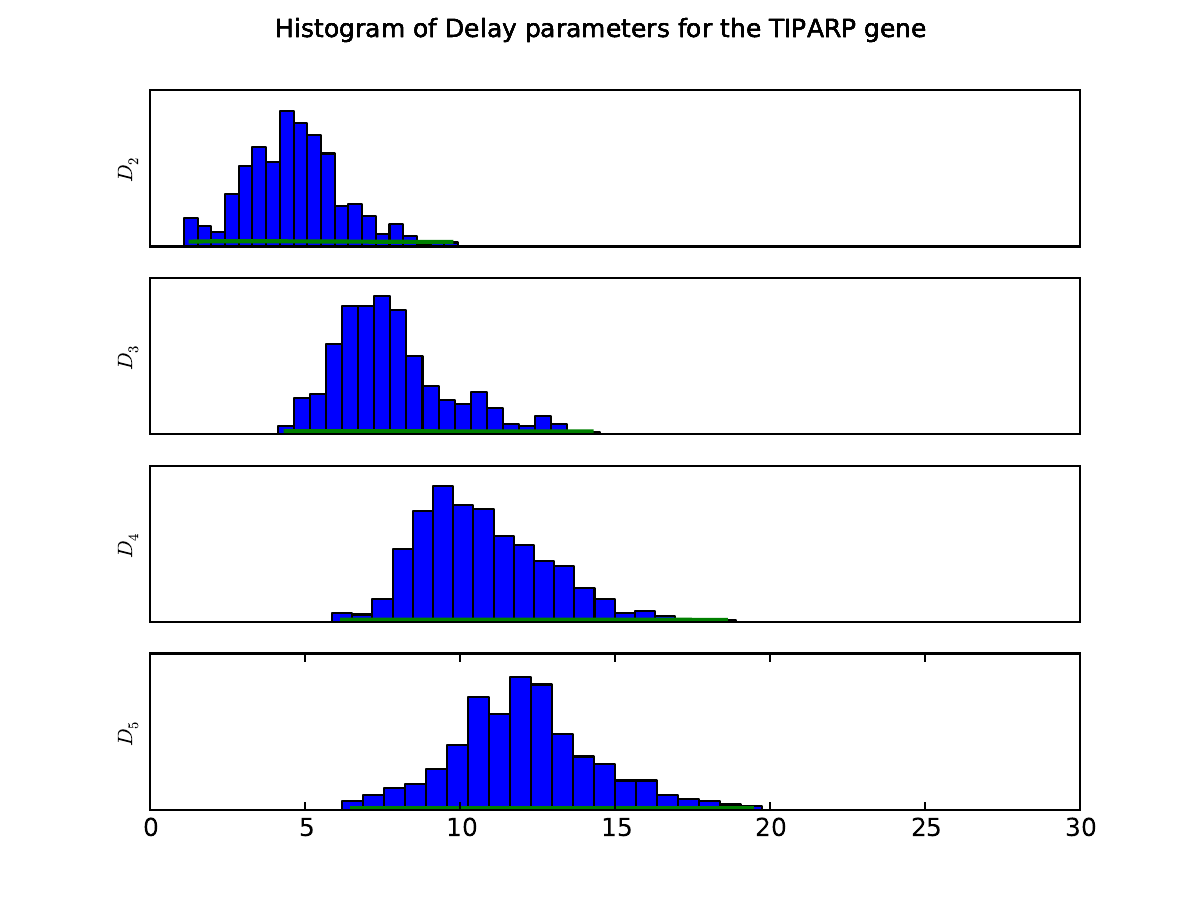}
\label{fig:palld_2}
}
\subfigure[]{
\includegraphics[width=0.38\textwidth]{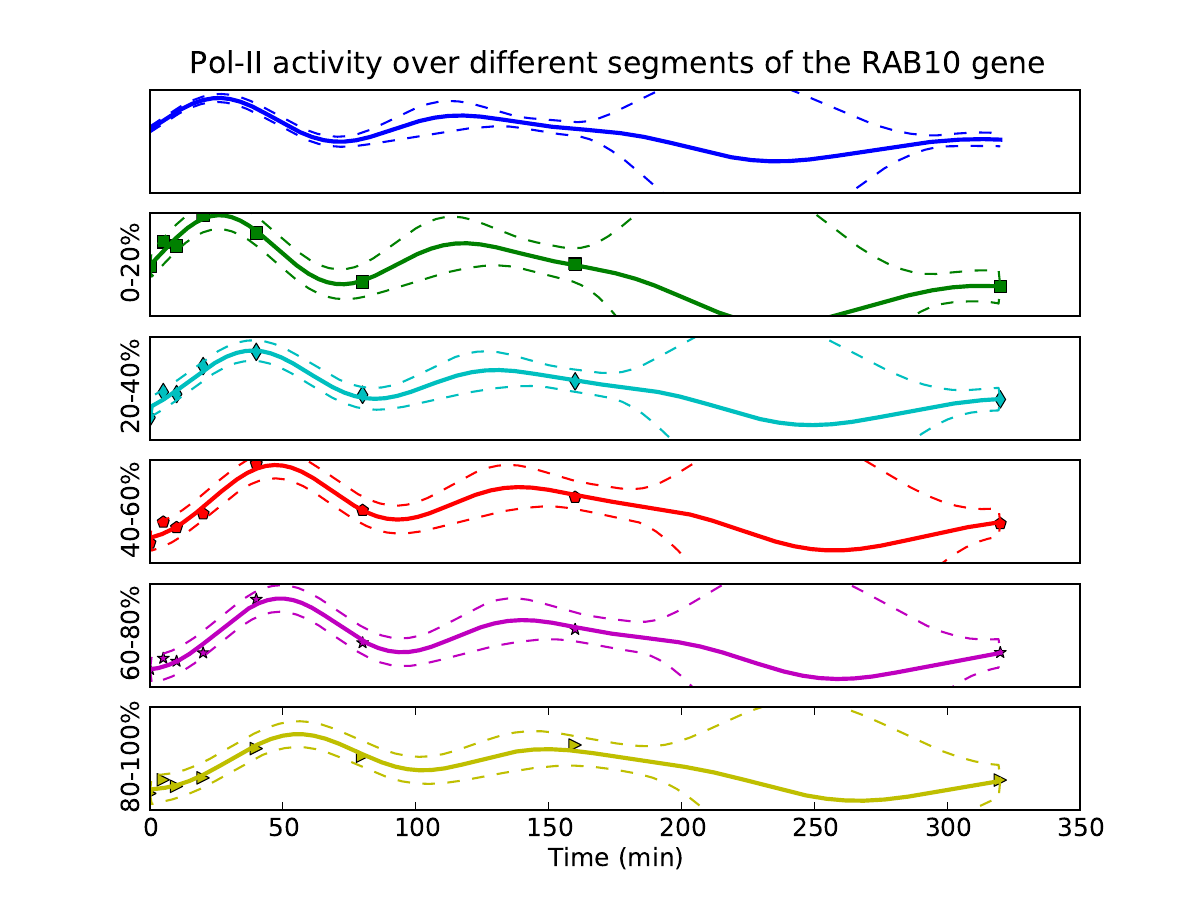}
\label{fig:siah2}
}
\subfigure[]{
\includegraphics[width=0.38\textwidth]{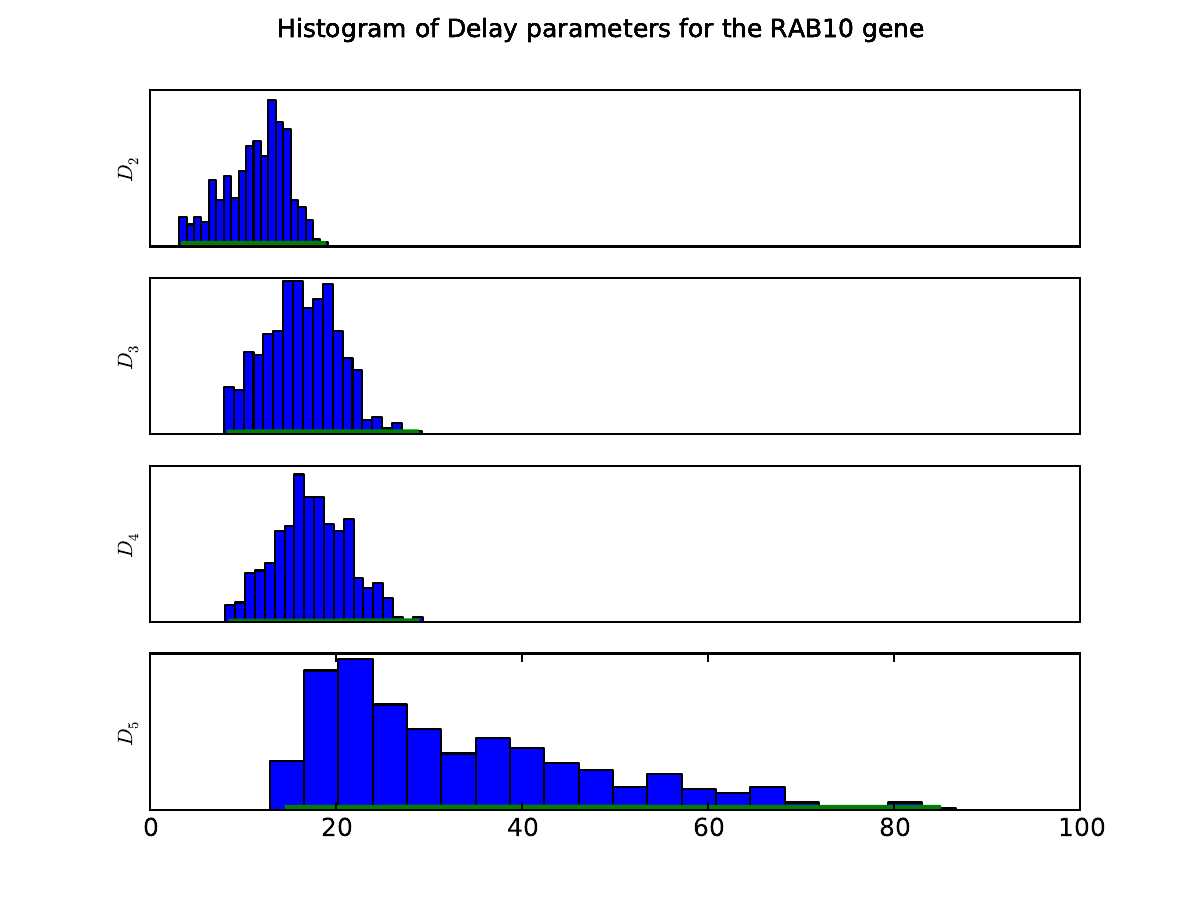}
\label{fig:siah2_2}
}
\caption[Optional caption for list of figures]{Inferred pol-II time profiles obtained for three of the top ten genes using ChIP-seq data. The top panel of each figure shows the inferred distribution of the latent funtion $f(t)$. The next five panels show the inferred profiles for the five gene segments corresponding to $0-20\%,\ldots,80\%-100\%$ of the gene. The figures on the right are the delay histograms}
\label{fig:twave_lg}
\end{figure}

Figure \ref{fig:twave_lg} shows the inferred pol-II time profile and histogram of the samples of the delay parameters for three of the top 10 genes found to fit the model well. 
We note that a relatively small number of activated genes fit the model well. This is primarily because for shorter genes the pol-II occupancy quickly rises over the whole gene such that the temporal resolution of the data cannot capture the wave as it traverses the gene body. With a closer or more evenly spaced time course we would expect a good fit for a greater proportion of activated genes. 

\begin{figure}[ht]
\centering
\subfigure[]{
\includegraphics[width=0.38\textwidth]{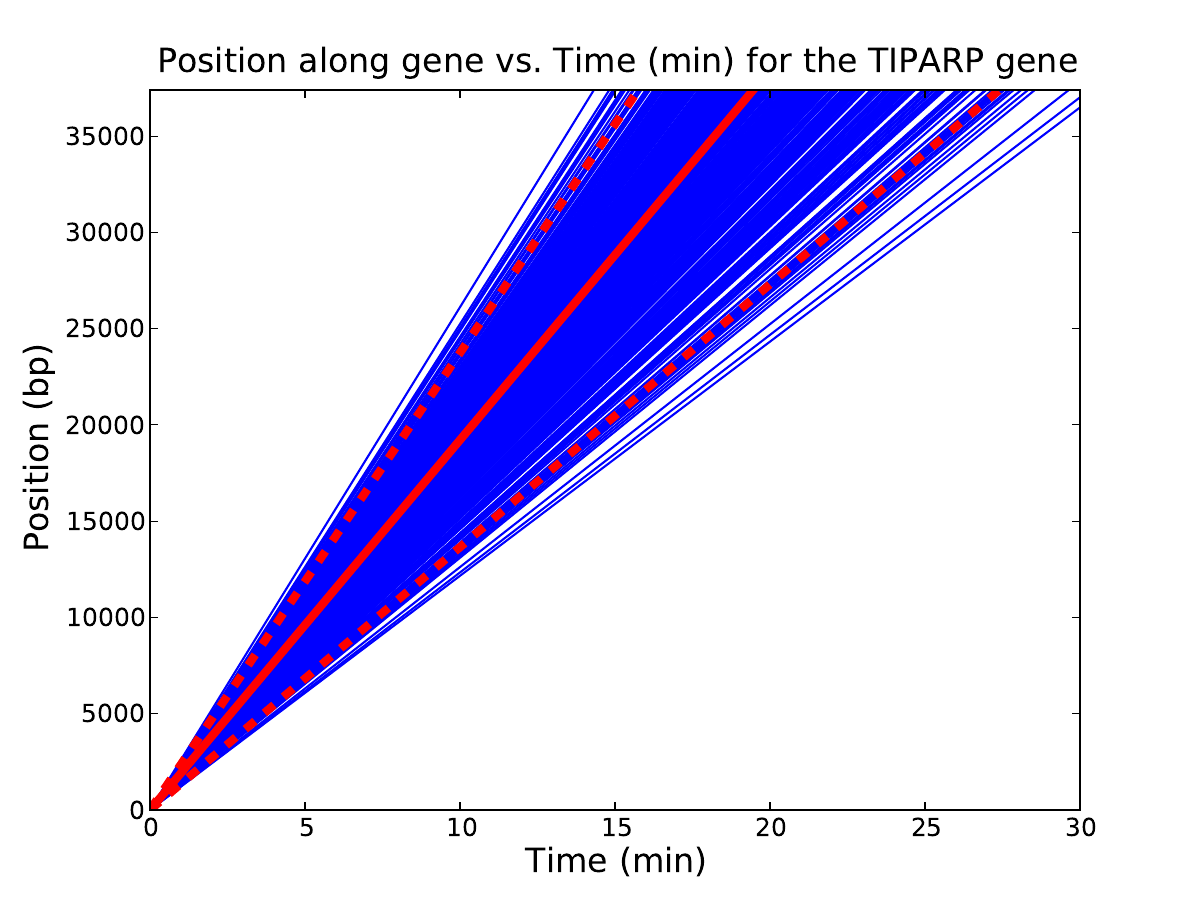}
\label{fig:tiparp_samps}
}
\subfigure[]{
\includegraphics[width=0.38\textwidth]{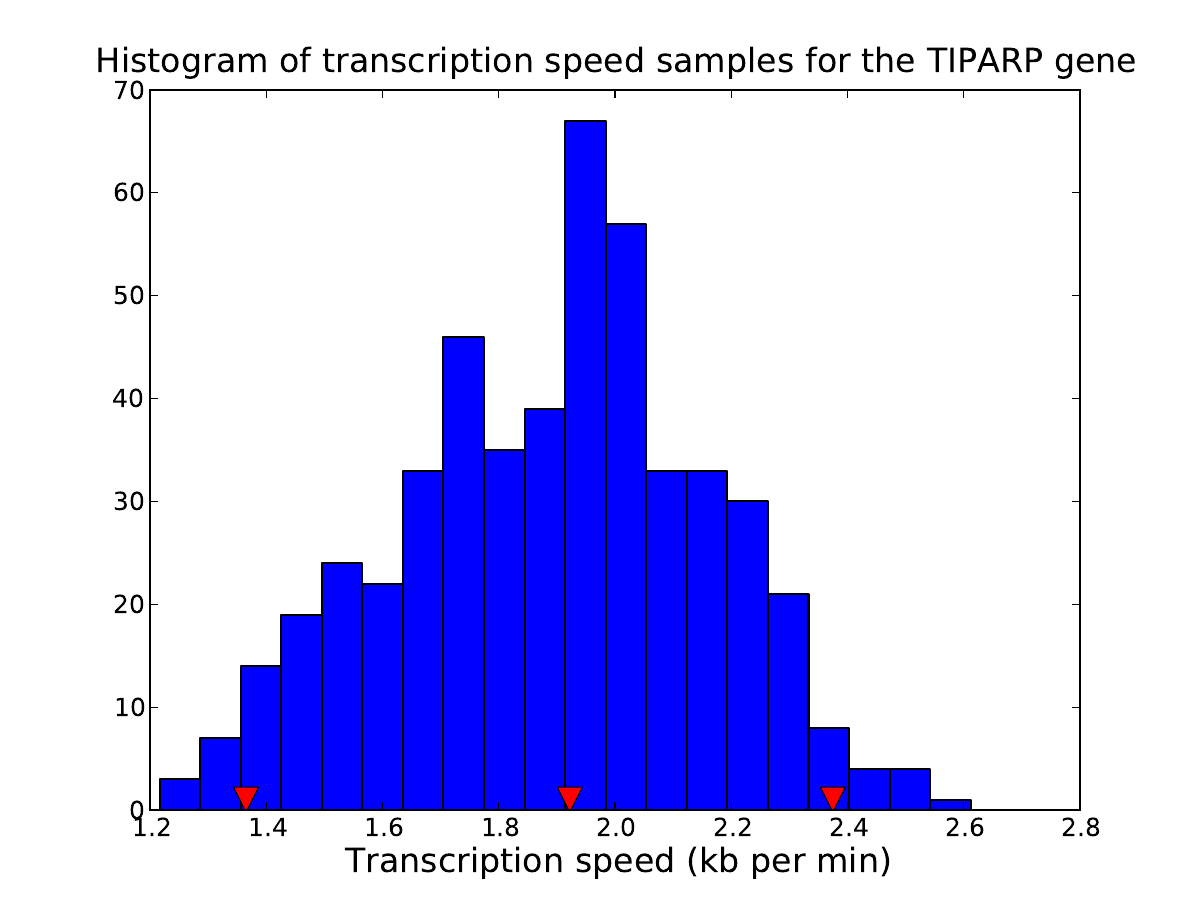}
\label{fig:tiparp_hist}
}
\label{fig:tiparp_speed_comp}
\caption[Optional caption for list of figures]{Linear regression plots using the delay samples for the \textit{TIPARP} gene (a) and the histogram of speed samples (b). The 95\% confidence interval is indicated in (a) by the dashed red lines with the median represented by the solid red line. In (b) the 95\% confidence interval is indicated  by the red triangle markers (cf. Table \ref{tab:chip_rank}).}
\end{figure}

\begin{figure}[ht!]
\centering
\includegraphics[width=0.75\textwidth]{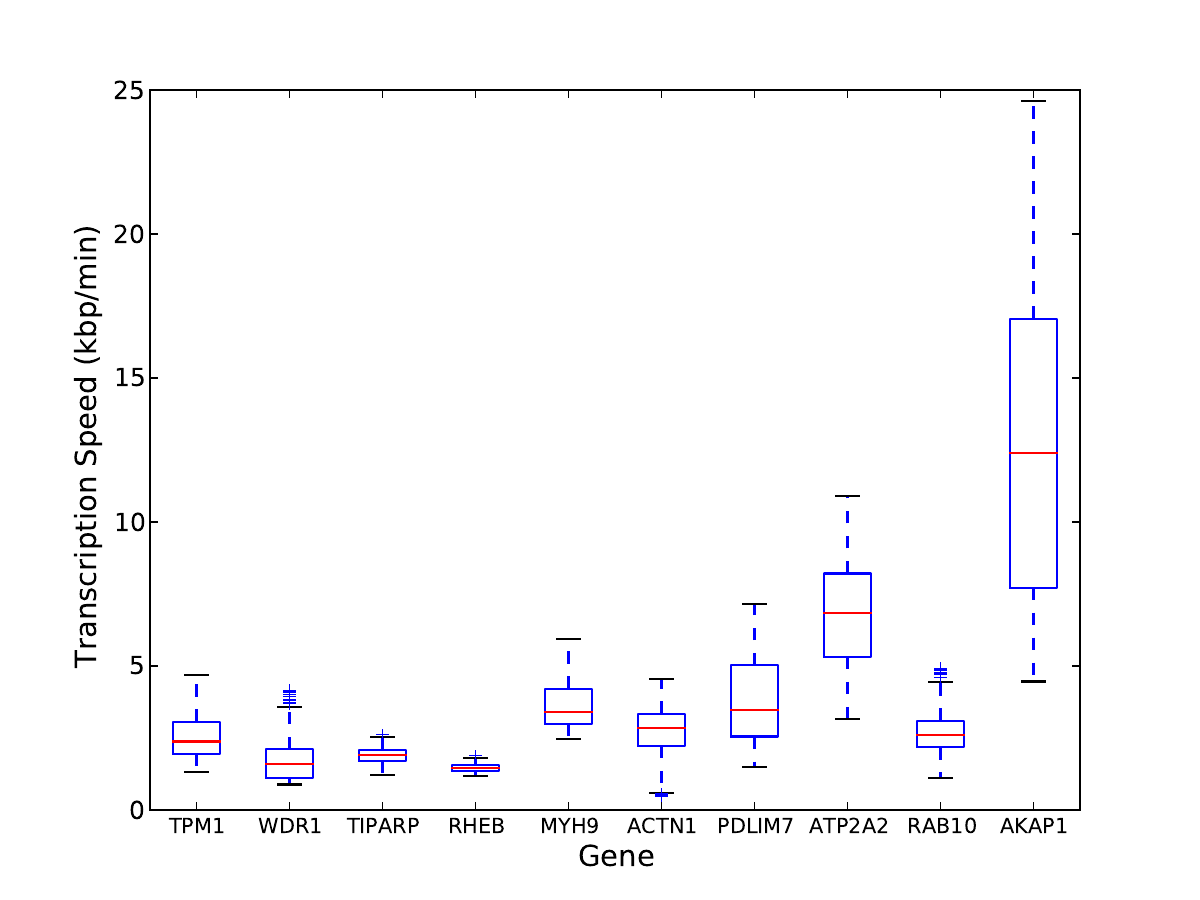}
\caption{Box plot of speed estimates for the top ten genes found to fit the transcription model well.}
\label{fig:speed_box}
\end{figure}

 Figure \ref{fig:tiparp_samps} shows the linear regression plots using the delay samples for the \textit{TIPARP} gene. Figure \ref{fig:tiparp_hist} shows the histogram of speed samples from which we can compute the confidence interval for the speed estimate. The 95\% confidence interval is indicated in Figure \ref{fig:tiparp_hist} by the red triangle markers (cf. Table \ref{tab:chip_rank}). Table \ref{tab:chip_rank} shows the average transcription speeds for the top 10 genes computed using the samples of the delay parameters. Figure \ref{fig:speed_box} shows a box plot of the average transcription speeds computed using the samples of the delay parameters for these genes. 

The advantage of fitting each of the delay parameters independently instead of enforcing a linear relationship is that it allows us to take into account phenomena such as pol-II pausing and provides a means to filter genes where the values of estimated delay are not naturally ordered.
Visual inspection of the inferred time series of the top ranked genes is consistent with a `transcription wave' traversing the gene. The transcription wave is especially evident in the longer genes \textit{MYH9} and \textit{RAB10}. This motivates a closer look at long genes. Table \ref{tab:chip_rank_lg} shows the average transcription speeds computed using the samples of the delay parameters for the 23 long genes found to fit the pol-II dynamics model well.
Grouping these genes according to the magnitude of the median transcription speed allows us to compare our results to those presented previously. From Table \ref{tab:chip_rank_lg} we see that 12 (52\%) of these genes have average transcription speeds between 2 and 4 kb per minute, a range that includes speeds previously reported in the literature \cite{Wada27102009,Singh_2009}.

\begin{center}
\begin{table}[ht!]
\begin{center}
\begin{tabular}{|c|c|c|c|c|} 
\hline 
Gene &Length (bp) &2.5\%&50\%& 97.5\%\\ 
\hline
TPM1 & 22196 & 1.6 & 2.4 & 4.1 \\
\hline
WDR1 & 42611 & 1.0 & 1.6 & 3.5 \\
\hline
TIPARP & 32353 & 1.4 & 1.9 & 2.4 \\
\hline
RHEB & 53913 & 1.2 & 1.5 & 1.7 \\
\hline
MYH9 & 106741 & 2.6 & 3.4 & 5.5 \\
\hline
ACTN1 & 105244 & 0.6 & 2.8 & 4.2 \\
\hline
PDLIM7 & 14208 & 1.7 & 3.5 & 6.4 \\
\hline
ATP2A2 & 69866 & 3.6 & 6.8 & 10.2 \\
\hline
RAB10 & 103595 & 1.4 & 2.6 & 4.4 \\
\hline
AKAP1 & 36158 & 5.0 & 12.4 & 21.4 \\
\hline
\end{tabular} 
\end{center}
\caption{Transcription speed in kilobases per minute for the top ten genes that fit the transcription model well. We use a Bayesian MCMC method for parameter estimation which provides the posterior distribution of the transcription speed. We show the 2.5\%, 50\% and 97.5\% percentiles of the posterior distribution.
}
\label{tab:chip_rank}
\end{table} 
\end{center}

\begin{center}
\begin{table}[ht!]
\begin{center}
\begin{tabular}{|c|c|c|c|c|} 
\hline 
Gene &Length (bp) &2.5\%&50\%& 97.5\%\\  
\hline
ACTN1 & 105244 & 0.6 & 2.8 & 4.2 \\
\hline
ADCY1 & 148590 & 2.8 & 9.7 & 43.6 \\
\hline
ARHGEF10L & 158041 & 2.8 & 5.4 & 8.5 \\
\hline
EPB41L1 & 120374 & 0.2 & 0.4 & 2.0 \\
\hline
EPS15L1 & 110355 & 16.1 & 30.0 & 43.1 \\
\hline
FARP1 & 102125 & 1.7 & 2.9 & 7.9 \\
\hline
FLNB & 163856 & 0.2 & 1.5 & 3.7 \\
\hline
ITPK1 & 179005 & 0.3 & 2.9 & 6.8 \\
\hline
JAK1 & 133282 & 0.6 & 2.2 & 4.2 \\
\hline
JAK2 & 142939 & 0.6 & 2.4 & 5.3 \\
\hline
KIAA0232 & 101441 & 0.9 & 2.3 & 4.0 \\
\hline
KIF21A & 150163 & 1.0 & 2.1 & 3.8 \\
\hline
LARP1 & 104702 & 0.7 & 2.0 & 3.8 \\
\hline
MYH9 & 106741 & 2.6 & 3.4 & 5.5 \\
\hline
NCOR2 & 243050 & 6.5 & 10.9 & 20.5 \\
\hline
NRIP1 & 103571 & 2.9 & 4.7 & 6.4 \\
\hline
PKIB & 116142 & 0.6 & 1.0 & 2.4 \\
\hline
RAB10 & 103595 & 1.4 & 2.6 & 4.4 \\
\hline
RAB31 & 154326 & 0.7 & 1.6 & 3.0 \\
\hline
RASA3 & 150902 & 0.6 & 1.4 & 6.0 \\
\hline
SHB & 153316 & 0.5 & 3.1 & 5.0 \\
\hline
WWC1 & 180244 & 1.9 & 3.6 & 5.6 \\
\hline
ZNF644 & 106174 & 0.1 & 0.2 & 1.5 \\
\hline
\end{tabular} 
\end{center}
\caption{Transcription speed in kilobases per minute for long genes between 100 and 300 kilobases long}
\label{tab:chip_rank_lg}
\end{table} 
\end{center}

\subsubsection*{Clustering of Promoter Activity Profiles}

The inferred latent functions for each gene model the pol-II activity adjacent to the promoter. Clustering these profiles and examining the average profiles of each cluster allows us to visualise the general trends and also classify genes according to the immediacy and nature of the response. This provides an alternative to clustering based on mRNA abundance data (from microarray or RNA-Seq experiments) which is regulated both by mRNA production and degradation processes. The production of mRNA may be delayed relative to the actual activation of transcription at the promoter causing genes which are actually triggered at the same time to show different rates of mRNA production. Differences in degradation rate can also influence mRNA abundance profiles. It may therefore be difficult to distinguish early and delayed transcriptional regulation from mRNA abundance data.

To classify the profiles we sample the mean of the latent function ($f(t)$ in equation \ref{eq:obs}) and use PUMA-CLUST \cite{puma_09} to cluster the genes. PUMA-CLUST has the advantage of taking into account the uncertainty of the latent function when clustering the profiles. This uncertainty is computed from the posterior covariance of $f(t)$.

\begin{figure}[ht!]
\centering
\includegraphics[width=0.75\textwidth]{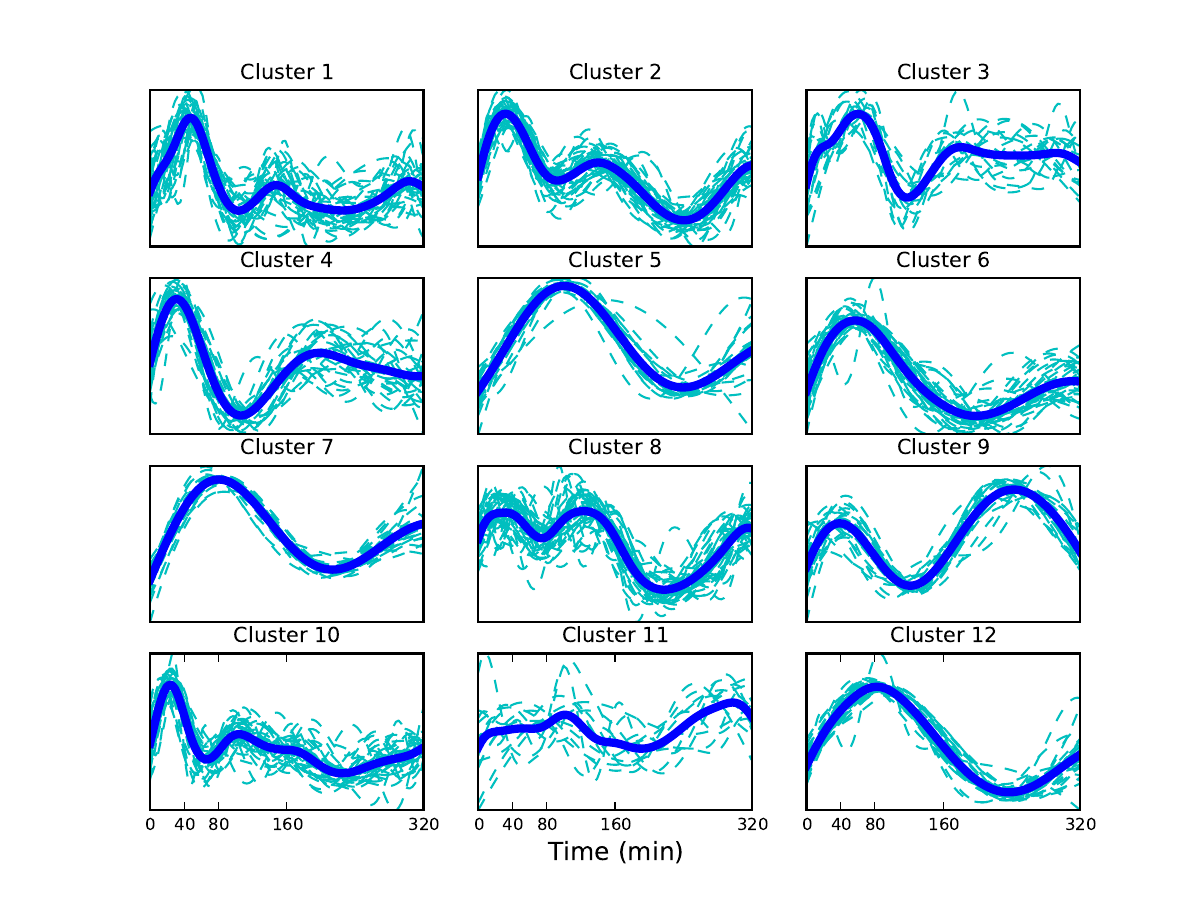}
\caption{Clusters of promoter activity profiles. The mean profile in each cluster is shown by the bold line.}
\label{fig:clusters_gw}
\end{figure}

The 383 genes found to fit the model well were grouped into 12 clusters (Figure \ref{fig:clusters_gw}) with the optimal number of clusters determined by the Bayesian Information Criterion. To determine the speed of the response in each cluster, we compute the peak time of the mean profile for each cluster (see Table \ref{tab:peak_time}). We used the Genomatix Pathway System (GePS) to look for enriched canonical pathways ($p$-value $<0.01$) in each cluster (supplementary material, Table \ref{tab:pathway_gw}) and performed a Gene Ontology (GO) analysis of the clusters using the DAVID tool \cite{DAVID_Huang_08,DAVID_Huang_09} (supplementary material, Tables \ref{tab:DAVID_GO_1}-\ref{tab:DAVID_GO_3}) showing that clusters are enriched for a number of different GO categories. The GO analysis identified early peaking clusters such as 2, 4 and 10 as enriched for nucleotide binding proteins consistent with many early genes being involved in downstream transcriptional regulation. The clustering of the pair of genes \textit{JAK1} and \textit{JAK2} in cluster 10, which has a prominent early peak, suggests that the response of both genes to E2 is rapid and coordinated. Since these genes are known to act together in several biological pathways such as the IL-6 signaling pathway and the IFN gamma signaling pathway, their appearance in the same cluster suggests that the clustering is likely to reveal other biologically significant relationships. 

\begin{figure}[ht!]
\centering
\subfigure[]{
\includegraphics[width=0.40\textwidth]{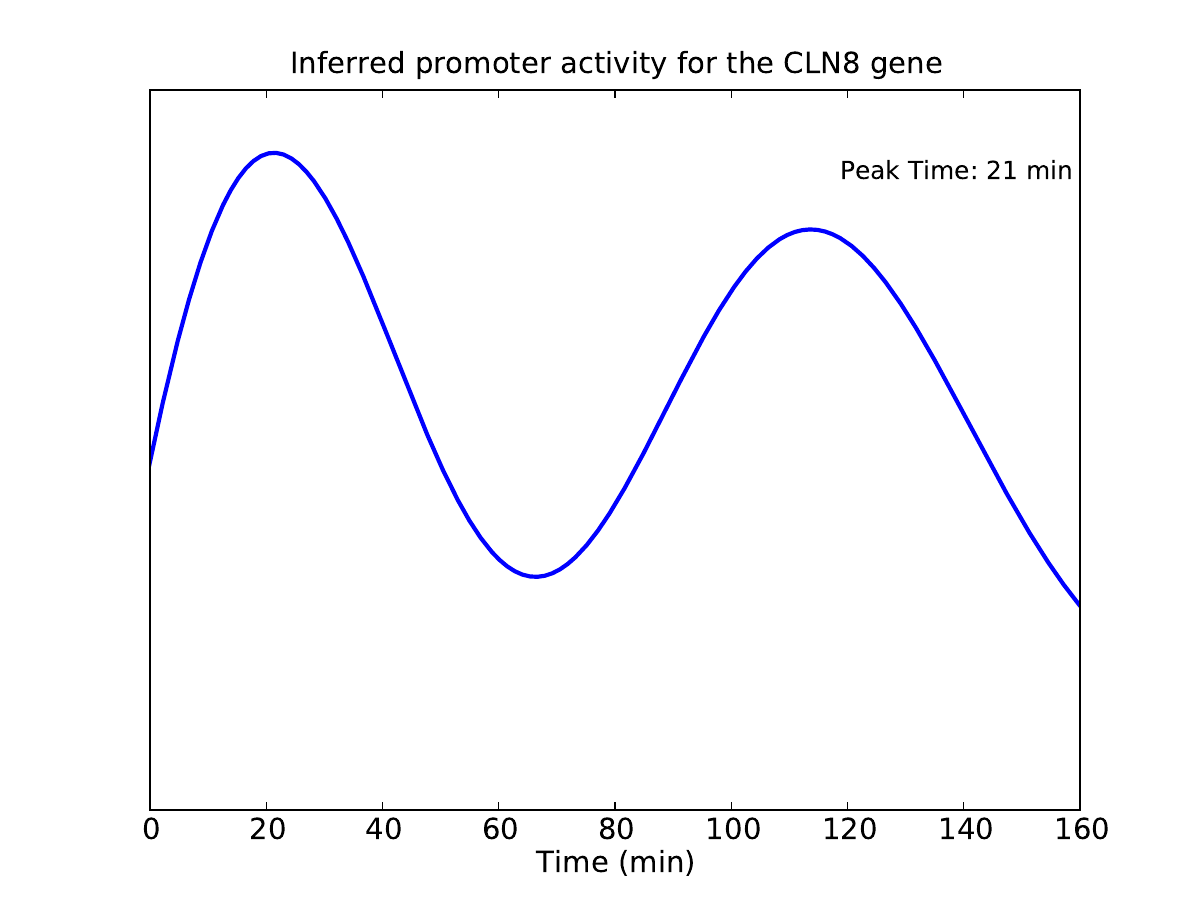}
\label{fig:cln8}
}
\subfigure[]{
\includegraphics[width=0.40\textwidth]{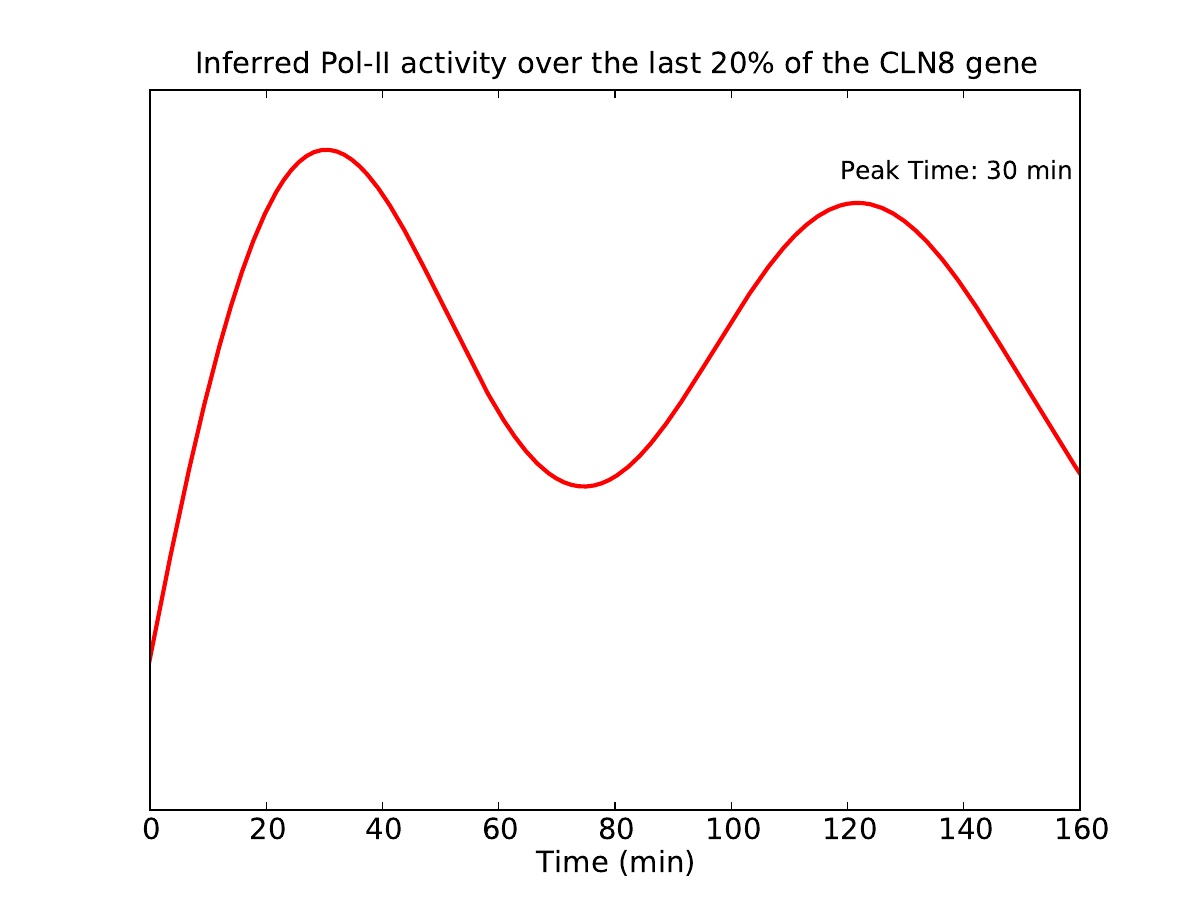}
\label{fig:cnl8_2}
}
\subfigure[]{
\includegraphics[width=0.40\textwidth]{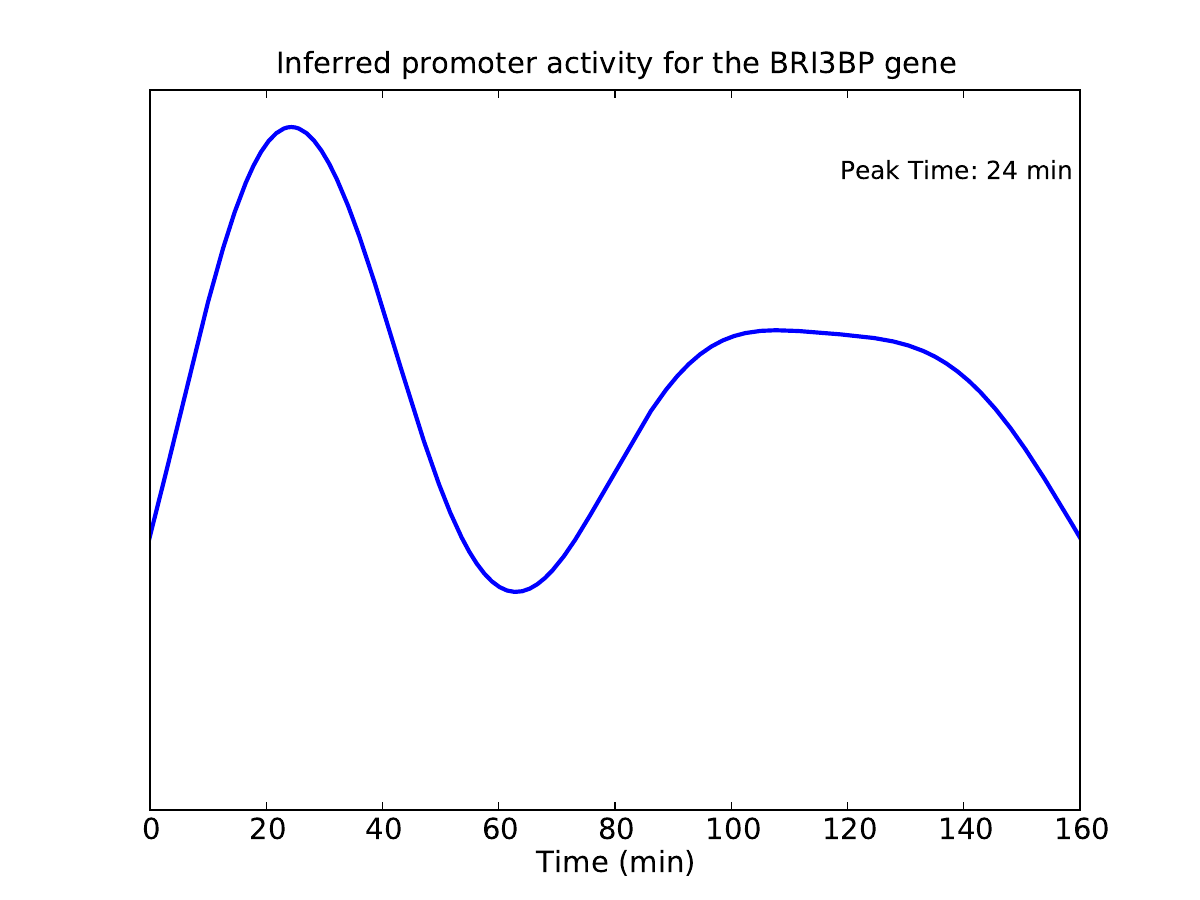}
\label{fig:bri3bp}
}
\subfigure[]{
\includegraphics[width=0.40\textwidth]{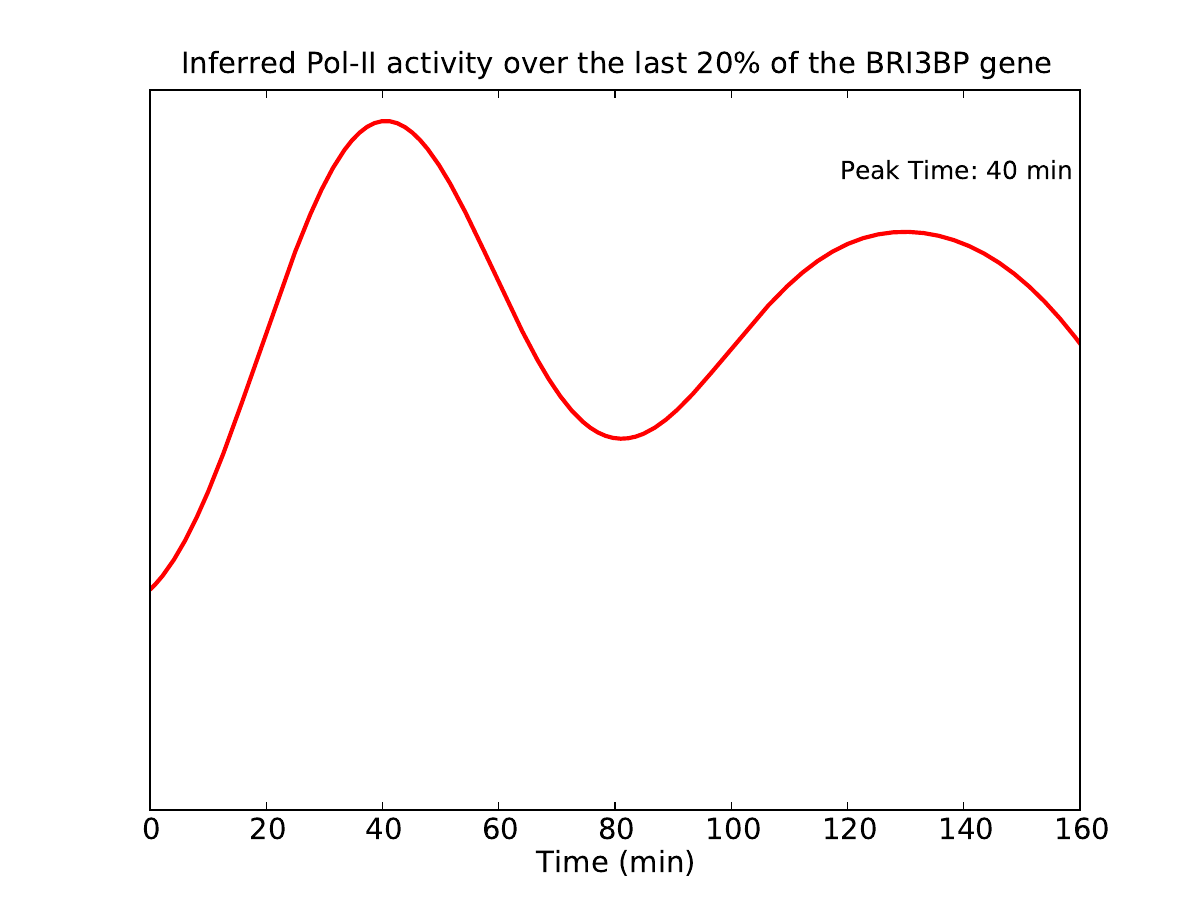}
\label{fig:bri3bp_2}
}
\subfigure[]{
\includegraphics[width=0.40\textwidth]{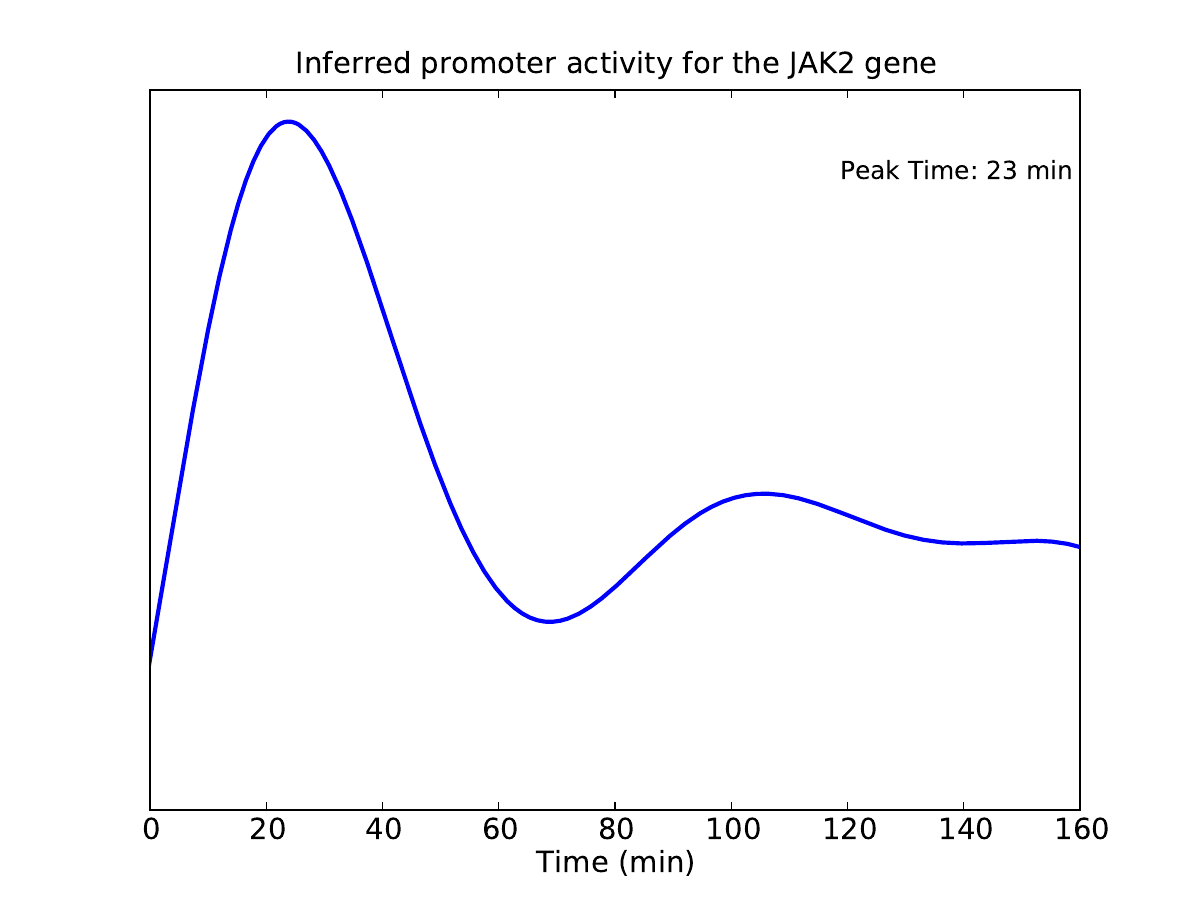}
\label{fig:jak2}
}
\subfigure[]{
\includegraphics[width=0.40\textwidth]{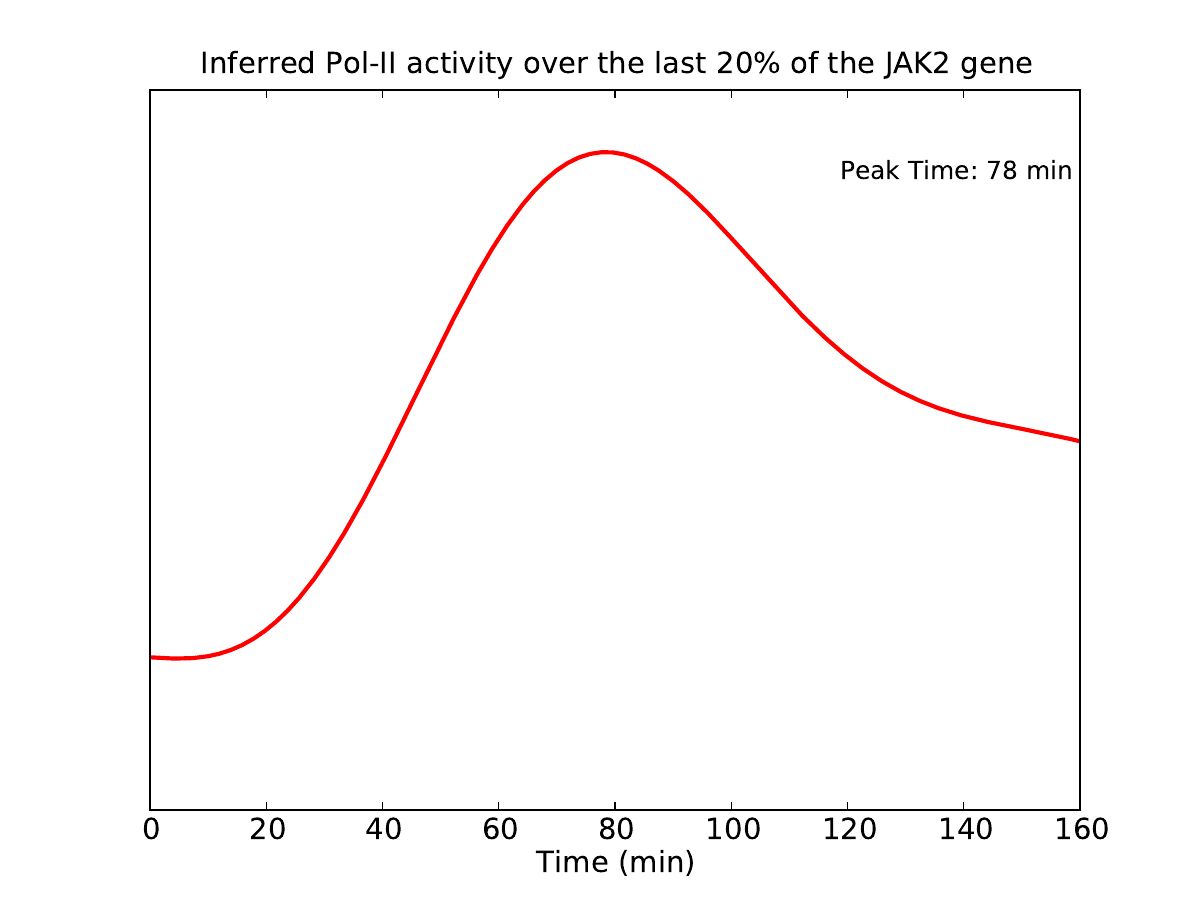}
\label{fig:jak2_2}
}
\caption[Optional caption for list of figures]{Inferred promoter profiles and pol-II activity over the final 20\% of the gene for three genes in cluster 10.}
\label{fig:latent_lastseg}
\end{figure}

\begin{figure}[ht!]
\centering
\subfigure[]{
\includegraphics[width=0.48\textwidth]{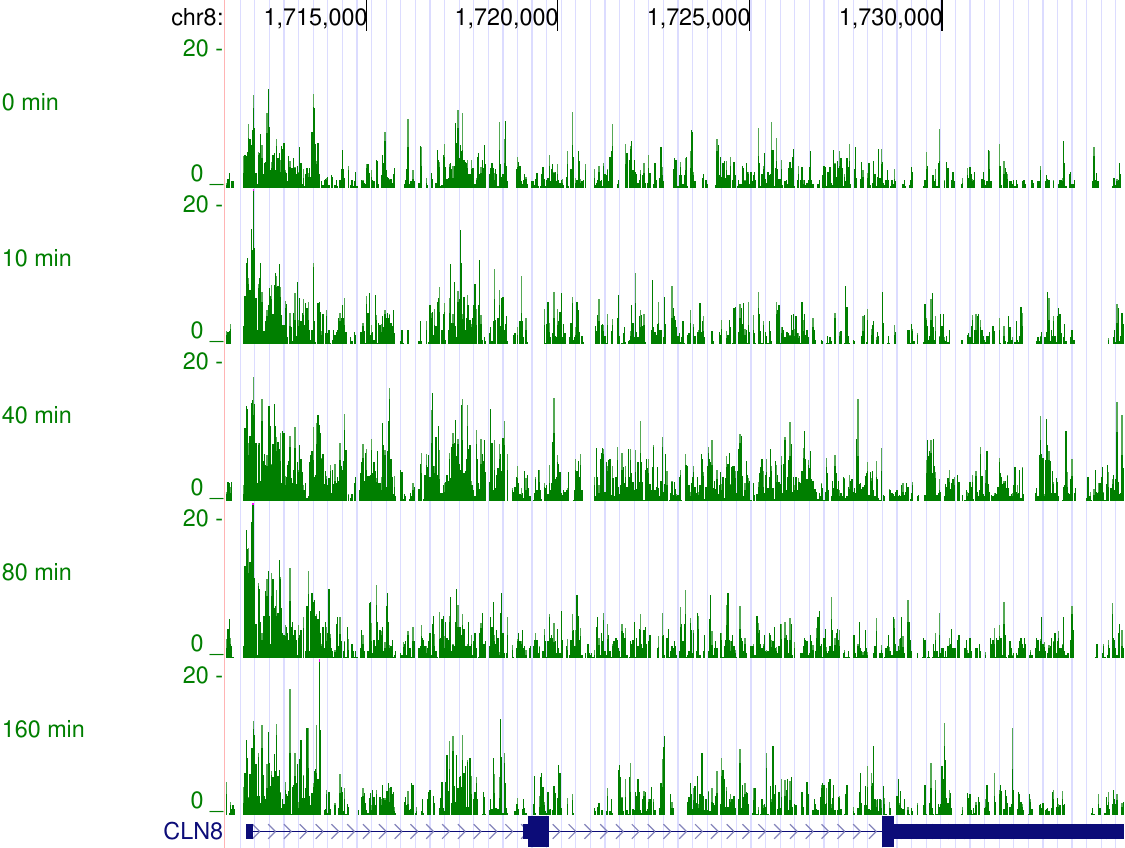}
\label{fig:cln8_chipseq}
}
\subfigure[]{
\includegraphics[width=0.48\textwidth]{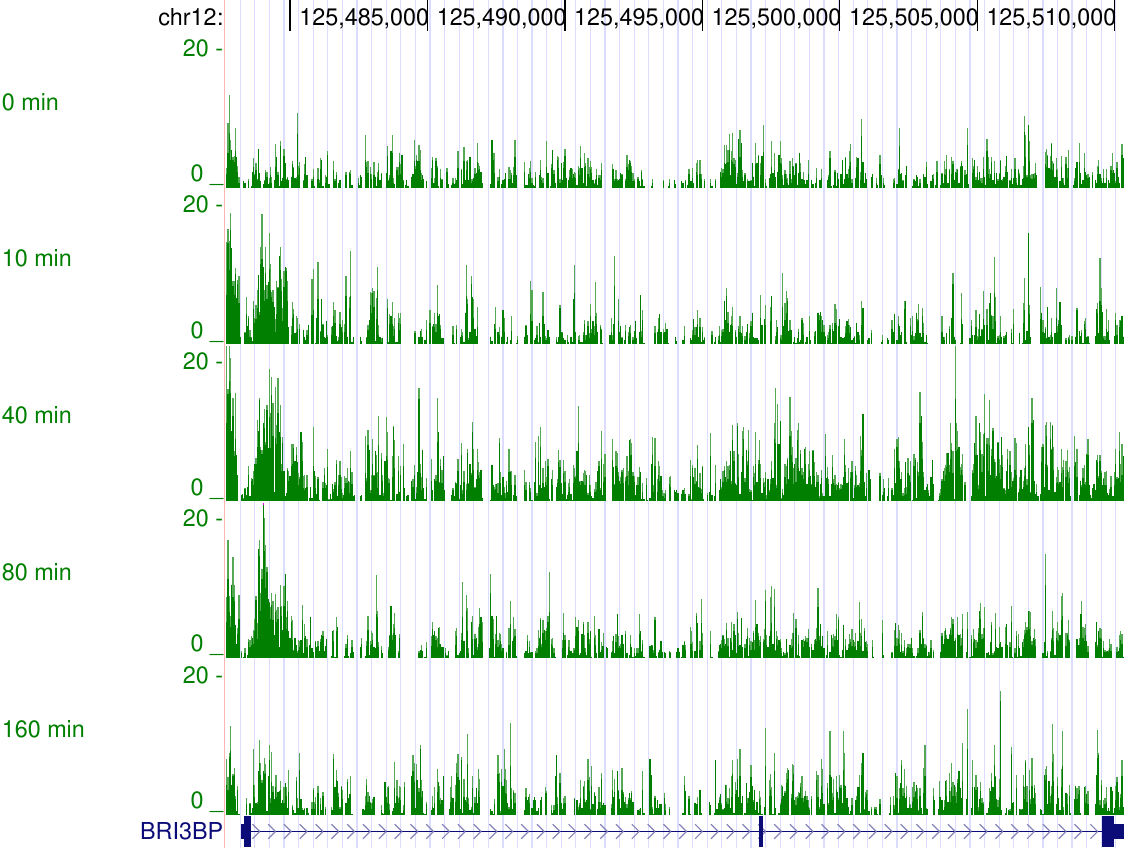}
\label{fig:bri3bp_chipseq}
}
\subfigure[]{
\includegraphics[width=0.48\textwidth]{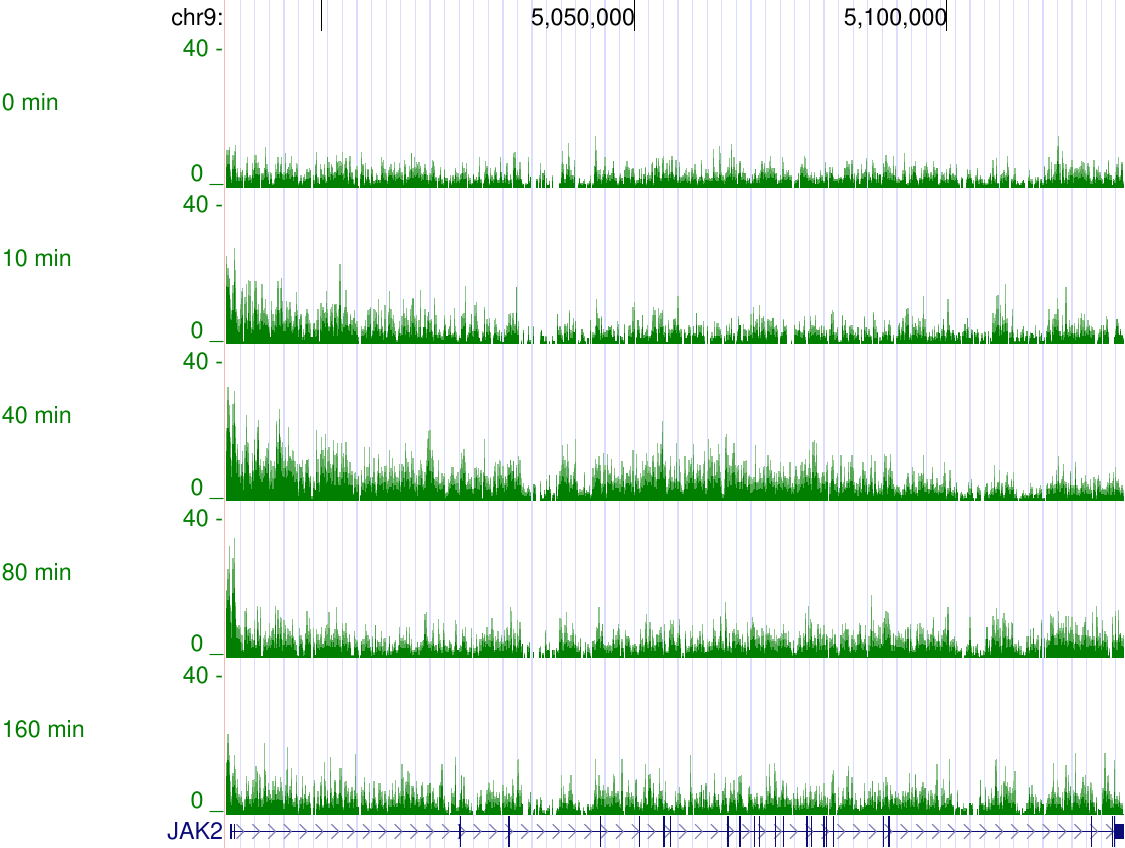}
\label{fig:jak2_chipseq}
}
\caption[Optional caption for list of figures]{ChIP-seq reads for three genes in cluster 10: \textit{CLN8}, \textit{BRI3BP} and \textit{JAK2}.
 }
\label{fig:raw_chipseq}
\end{figure}

A closer look at the inferred pol-II promoter profiles of some examples in cluster 10, the earliest peaking cluster, and the corresponding inferred pol-II profiles over the last 20\% of the genes reveals the possible influence of gene length on mRNA production and how clustering the inferred promoter profiles can account for this influence and uncover potential co-regulation. Figure \ref{fig:latent_lastseg} shows the inferred promoter profiles and the inferred pol-II profiles over the last 20\% for three genes \textit{CLN8}, \textit{BRI3BP} and \textit{JAK2} in cluster 10. Figure \ref{fig:raw_chipseq} shows the corresponding raw ChIP-seq reads. The lengths of the genes to the nearest kilobase are 23, 32 and 143 kb respectively. We see that despite the last segment profiles peaking at different times, the promoter profiles peak at approximately the same time. The difference in peak time over the final segment of the gene is most likely due to the length of the genes and accounts for the amount of time the pol-II takes to move down the gene. Such differences would mask potential co-regulation if we attempted to cluster genes based on their mRNA profiles. 

In Hah \textit{et al.} ~\cite{Hah_2011} GRO-seq was used to measure pol-II occupancy genome-wide when MCF-7 cells are treated
with estradiol (E2) at four time points (0, 10, 40 and 160 min after E2 treatment). In addition,
steady state levels of mRNA for 54 genes were measured using RT-qPCR at five time points (0, 10, 40, 160 and 320 min after E2 treatment). These data show a delay of between 1-3hr between peaks in the pol-II occupancy at the 5' end of a gene and peaks in the mRNA steady state ~\cite[Figure S4]{Hah_2011}. These data include the mRNA measurement for 20 genes whose corresponding GRO-seq data peak is at 40 minutes after E2 treatment. Six of these genes namely \textit{CASP7,
FHL2,
GREB1,
ITPK1,
NRIP1,
WWC1}
 are found to fit our pol-II model well with ChIP-seq data. Table \ref{tab:groseq_comp1} shows the peak time of the inferred promoter profile $T_p$, the peak time of the inferred pol-II profile over the last 20\% of the gene $T_{\mathrm{last}}$, the GRO-seq peak time as well as the mRNA peak time. For the GRO-seq and mRNA peak times we show the peak times from Hah \textit{et al.}~\cite[Figure S4]{Hah_2011} which are limited to the finite set of sampling times. We see that all mRNA peaks occur after $T_{\mathrm{last}}$. The large value of $T_{\mathrm{last}}$ for \textit{WWC1} which is a long gene $\sim 180$ kb in length corresponds to a late peak in mRNA at 320 minutes. This shows that the parameters obtained by our model are biologically plausible. Based solely on the GRO-seq data these genes were grouped together in ~\cite{Hah_2011} since they show a peak at 40min. However our modeling reveals a greater diversity in the nature of responses. In fact the six genes appear in three different early response promoter profile clusters (see Table  \ref{tab:groseq_comp1}).

In the supplementary material, we compare the clustering obtained from the inferred promoter profiles to that obtained if the time series of the raw ChIP-seq reads are clustered and show that our model has the potential to uncover relationships which may be missed if we only consider the raw ChIP-seq reads.

\begin{center}
\begin{table}[ht!]
\begin{center}
\begin{tabular}{|c|c|} 
\hline 
Cluster &Peak Time (min)\\ 
\hline
1 & 48 \\
\hline
2 & 32 \\
\hline
3 & 61 \\
\hline
4 & 32 \\
\hline
5 & 100 \\
\hline
6 & 58 \\
\hline
7 & 80 \\
\hline
8 & 122 \\
\hline
9 & 242 \\
\hline
10 & 22 \\
\hline
11 & 297 \\
\hline
12 & 80 \\
\hline
\end{tabular} 
\end{center}
\caption{Cluster peak time
}
\label{tab:peak_time}
\end{table} 
\end{center}

\begin{center}
\begin{table}[ht!]
\begin{center}
\begin{tabular}{|c|c|c|c|c|c|} 
\hline 
Gene&Cluster &$T_p$&$T_{20}$&GRO-seq Peak&mRNA Peak\\ 
\hline
CASP7&1&36& 47&40&160\\
\hline
FHL2&1&42&55&40&160\\
\hline
GREB1&2&30&46&40&320\\
\hline
ITPK1&2&36&64&40&160\\
\hline
NRIP1&10&22&40&40&160\\
\hline
WWC1&10&23&81&40&320\\
\hline
\end{tabular} 
\end{center}
\caption{The peak time of the inferred promoter profile $T_p$, the peak time of the inferred pol-II profile over the last 20\% of the gene $T_{20}$, the GRO-seq peak time as well as the mRNA peak time (from ~\cite[Figure S4]{Hah_2011}).}
\label{tab:groseq_comp1}
\end{table} 
\end{center}

\subsubsection*{Transcription factor binding}

We investigated the TF peaks in a 40 kbp region around the gene transcription start site for all genes in each cluster using ChIP-seq data for a number of TFs measured under similar experimental conditions (i.e. MCF-7 breast cancer cells treated with E2) in the cistrome database (\url{http://cistrome.org}). In earlier work on the estrogen interactome, Fullwood \textit{et al.} \cite{Fullwood_2009} suggest that most long range interactions between TF binding sites and gene enhancers are limited to a range of about 20kb. We therefore investigate the region from -20kb to 20kb relative to the TSS (results for other regions around the TSS ranging from 1 to 100 kb are shown in the supplementary material (Tables \ref{tab:tf_gw1} -\ref{tab:tf_gw5})).  Table \ref{tab:tf_gw} shows the number of genes with TF binding peaks for each cluster for 7 TFs namely ER$\alpha$ \cite{Welboren_2009}, FoxA1 \cite{Lupien2008}, c-Fos \cite{Joseph2010}, c-Jun \cite{Joseph2010}, c-MYC \cite{Hua_2009}, SRC-3 \cite{Lanz01042010}, TRIM24 \cite{Tsai2010}. We found that the TFs RAD21 \cite{Schmidt2010}, CTCF \cite{Schmidt2010} and  STAG1 \cite{Schmidt2010} are ubiquitously bound and not useful in uncovering cluster-specific TF binding. We investigate the statistical significance of the proportions of genes in each cluster with TF peaks in a 40kb neighborhood of the TSS by comparing the observed proportions to those we would expect in clusters of the same size drawn at random from the set of all genes. In Table \ref{tab:tf_gw} statistically significant ($p$-value $<0.05$) proportions are indicated in red (larger than expected) and green (lower than expected). For $p$-values less than $0.01$, the associated $p$-values are indicated in parentheses according to the following scale (***: $p<0.0001$,**: $p<0.001$,*:$p<0.01$).

Interestingly, clusters 1, 2, 4, and 10, which show an early peak in the mean promoter profile, are
all enriched for ER$\alpha$ and FOXA1. These clusters, with the exception of cluster 4, were also found to be enriched for the ER$\alpha$ motif near the promoter. The enrichment of both ER$\alpha$ and FOXA1 in these clusters is in line with conclusions drawn in Hurtado \textit{et al.} \cite{Hurtado_2011} where it was suggested FOXA1 mediates ER$\alpha$ binding. We also investigated the overlap of the binding sites for ER$\alpha$ and FOXA1 both in the 151 genes belonging to these clusters and genome-wide using the peaks obtained from \cite{Welboren_2009} (ER$\alpha$) and \cite{Lupien2008} (FOXA1) and reported in the cistrome database. We investigated the 40kb region -20kbp to 20kbp relative to the TSS. Table \ref{tab:ER_FOXA1} shows the number of ER$\alpha$ and FOXA1 peaks and the overlap (Two peaks are said to overlap if they have at least one base pair in common). We see that when we consider the rapid response genes in clusters 1, 2, 4, and 10 the percentage of overlap increases to 16\% (35/220) whereas the overlap is 9\% (956/11056) when we consider all genes. The significance associated with this elevated overlap is p=0.004 given the null hypothesis of a random gene list of the same size (results for other regions around the TSS ranging from 1 to 100 kb are shown in the supplementary material (Tables \ref{tab:ER_FOXA1_1} -\ref{tab:ER_FOXA1_4}))). 
 Taken together, the results in Tables \ref{tab:tf_gw} and \ref{tab:ER_FOXA1} identify genes that respond to E2, with clusters 1, 2, 4 and 10 most likely to contain the earliest estrogen responsive genes. 

\begin{center}
\begin{table}[ht!]
\begin{center}
\begin{tabular}{|c|c|c|c|c|c|c|c|} 
\hline 
Cluster&\multicolumn{7}{|c|}{TFs}\\ \cline{2-8}
&ER$\alpha$&FOXA1&c-FOS&c-JUN&MYC&SRC-3&TRIM24\\ 
\hline
1 (37)&{\color{red} 27 }(**)& {\color{red} 14 }& {\color{red} 16 }(*)& 6& 4& {\color{red} 25 }(*)& 27 \\
\hline
2 (47)&{\color{red} 31 }(*)& {\color{red} 19 }(*)& {\color{red} 16 }& 7& {\color{red} 7 }& {\color{red} 36 }(***)& {\color{red} 38 } \\
\hline
3 (18)&11& 5& 7& {\color{red} 5 }& {\color{red} 6 }(**)& 11& 12 \\
\hline
4 (29)&{\color{red} 20 }(*)& {\color{red} 11 }& 9& {\color{red} 7 }& 2& 18& 23 \\
\hline
5 (27)&15& 4& 6& {\color{red} 8 }(*)& {\color{red} 9 }(***)& 16& 19 \\
\hline
6 (40)&{\color{red} 27 }(*)& 8& 12& 7& 4& {\color{red} 25 }& 31 \\
\hline
7 (24)&10& 6& 5& {\color{red} 6 }& 3& 13& 19 \\
\hline
8 (47)&{\color{red} 32 }(*)& 10& 14& {\color{red} 14 }(**)& {\color{red} 8 }& {\color{red} 31 }(*)& {\color{red} 40 }(*) \\
\hline
9 (26)&{\color{red} 18 }& 7& {\color{red} 11 }(*)& {\color{red} 11 }(***)& 3& 12& {\color{red} 22 } \\
\hline
10 (38)&{\color{red} 30 }(***)& {\color{red} 14 }& {\color{red} 15 }(*)& 2& 1& {\color{red} 29 }(**)& {\color{red} 32 }(*) \\
\hline
11 (13)&5& 2& {\color{red} 7 }(*)& {\color{red} 4 }& 2& 7& {\color{red} 13 }(*) \\
\hline
12 (37)&19& 8& 12& {\color{red} 11 }(**)& 4& {\color{red} 23 }& 29 \\
\hline
\end{tabular} 
\end{center}
\caption{Analysis of transcription factor binding in 40kbp regions of genes in gene clusters obtained from inferred promoter activity profiles. The number in parentheses in the first column is the cluster size. For each TF, we show the number of genes with peaks. Statistically significant proportions ($p$-value $<0.05$) are indicated in red (larger than expected). For $p$-values less than $0.01$, the associated $p$-values are indicated in parentheses according to the following scale (***: $p<0.0001$,**: $p<0.001$,*:$p<0.01$).
}
\label{tab:tf_gw}
\end{table} 
\end{center}

\begin{center}
\begin{table}[ht!]
\begin{center}
\begin{tabular}{|c|c|c|c|} 
\hline 
Genes&\# of ER$\alpha$ peaks& \# of FOXA1 peaks &ER$\alpha$ and FOXA1 overlap\\ 
\hline
Clusters 1, 2, 4, and 10 (151)&220 (112) &86 (44) &35 (0.004)\\
\hline
All genes ($\sim$ 20,000)&11056&4626&956 \\
\hline
\end{tabular} 
\end{center}
\caption{Overlap of ER$\alpha$ and FOXA1 binding in a 40 kb region around the TSS. The numbers in parentheses in the first column are the number of genes. In each TF peak column, we show the expected number of peaks in a set of random random genes of the same size in parentheses. In the overlap column the associated p-value is shown in parentheses.}
\label{tab:ER_FOXA1}
\end{table} 
\end{center}

\section*{Discussion}
In this work we have presented a methodology for modelling transcription dynamics and employed it to determine the transcriptional response of breast cancer cells to estradiol. To capture the movement of pol-II down the gene body, we model the observed pol-II occupancy time profiles over different gene segments as the delayed response of linear systems to the same input. The input is assumed to be drawn from a Gaussian process which models the pol-II activity adjacent to the gene promoter. Given observations from high-throughput data such as pol-II ChIP-Seq data, we are able to infer this input function and estimate the pol-II activity at the promoter. This allows us to differentiate transcriptionally engaged pol-II from pol-II paused at the promoter and yields good estimates of transcriptional activity.

In addition to estimating the transcriptional activity at the promoter, inferring the pol-II occupancy time profiles over different gene segments allows us to compute the transcription speed. We expect the delay parameters of different gene segments to be non-decreasing and this provides a natural way to determine genes that are being actively transcribed in response to E2.

Clustering the inferred promoter activity profiles allows us to investigate the nature of the response and  group genes that are likely to be co-regulated. We found that the four clusters significantly enriched for both ER$\alpha$ and FOXA1 binding within 40kb according to public ChIP-Seq data were those that showed the earliest peak in pol-II activity at the promoter. ER$\alpha$ and FOXA1 ChIP peaks in the neighbourhood of these genes were also more likely to be overlapping than the average for ChIP-identified binding events of these TFs genome-wide. This observation provides some support for the previously proposed role of FOXA1 as a mediator of early transcriptional response in estrogen signalling. These results also show that our method can help regulatory network inference. The inferred promoter activity profiles pinpoint the times of transcriptional activation very accurately without confounding transcriptional delays. As genes with similar inferred promoter activity profiles are likely to have similar TF binding profiles, they are likely to be co-regulated as well. The promoter profiles should therefore lead to more accurate predictions of regulator-target relationships using time-course-based methods (e.g. \cite{Honkela_2010}) than using expression time course data.

As well as modelling transcriptional speed and transcriptional activity profiles, the proposed modelling approach may have other useful applications. For example, recent research has uncovered a link between transcription dynamics and alternative splicing \cite{Shukla_2011}. It is believed that aberrant splicing can cause disease and a number of studies have tried to understand the mechanisms of alternative splicing \cite{Tazi_2009}. The proposed model can potentially be used to identify transcriptional pausing events, and such results could be usefully combined with inference of splice variation from RNA-Seq datasets from the same system. Also, with the increasing availability of high-throughput sequencing data exploring multiple layered views of the transcription process and its regulation, the convolved modelling approach developed here has the potential to be usefully applied to more complex coupled spatio-temporal datasets.

\section*{Acknowledgments}
We thank Nancy Bretschneider for running the mappings to generate the bed-files for this publication. We thank Dr. Jarnail Singh, and Dr. Richard A. Padgett for making data from their paper available.


\clearpage
\section*{Supplementary information}

\subsection*{Priors}
The parameters $\Theta=\{\sigma_f,\ell_f,\{\alpha_i,D_i,\ell_i,\sigma_i\}_{i=1}^I\}$ are positive and bounded.
In the experiments we use the bounds shown in Table \ref{tab:bounds} with $D_1$ fixed at zero, $\sigma_f=1$ and the values $\sigma_i$ tied to single value. To determine the delay bounds, we assume that the value of $D_i$ is an indicator of how long it takes the `transcription wave' to reach the corresponding gene segment. That is $D_2$ is the amount of time it takes to transcribe 20\% of the gene, $D_3$ 40\% etc. We obtain the length $L$ of the gene from the hg19 annotation and use values of maximum and minimum expected speed ($s_{min}$ and $s_{max}$ respectively) to compute the delay bound. For example
\begin{displaymath}
D_2^{min}=\frac{0.2L}{s_{max}}\quad \mathrm{and} \quad D_2^{max}=\frac{0.2L}{s_{min}}
\end{displaymath}  
We use $s_{min}=50$ bp min$^{-1}$ and $s_{max} = 50$ kbp min$^{-1}$. These large bounds allow unbiased estimation of transcription speed. (Recent work on individual cells suggests speeds as high as 50kb per minute are possible \cite{Maiuri_11}.)

 We transform the parameters using a logit transform and work with unconstrained variables. For a parameter $\theta\in\Theta$ with corresponding minimum and maximum bounds $\theta_{min}$ and $\theta_{max}$ respectively we compute the transfromed variable $\gamma$ 
\begin{equation}
\label{eqn:sig}
\gamma=\log\Big(\frac{\theta-\theta_{min}}{\theta_{max}-\theta}\Big).
\end{equation}
  We place a Gaussian prior over the parameters in the transformed domain and draw samples from the posterior using the Hamiltonian Monte Carlo (HMC) algorithm \cite{Neal_HMC}. We have
\begin{equation}
\gamma\sim \mathcal{N}(\gamma|0,\sigma_\gamma).
\end{equation}
With $,\sigma_\gamma=2$ we obtain an approximately uniform prior in the untransformed domain yielding an uninformative prior.

\begin{center}
\begin{table}[ht!]
\begin{center}
\begin{tabular}{|c|c|c|} 
\hline 
Parameter&Minimum&Maximum\\ 
\hline
$\ell_f$&5 min&320 min\\
\hline
$\alpha_i$&0&100\\
\hline
$D_i$&$\frac{0.2(i-1)L}{s_{max}}$ min &$\frac{0.2(i-1)L}{s_{min}}$ min\\
\hline
$\ell_f$&5 min&320 min\\
\hline
$\sigma_i$&0&100\\
\hline
\end{tabular} 
\end{center}
\caption{Parameter bounds.}
\label{tab:bounds}
\end{table} 
\end{center}

To initialise the parameters for gradient optimisation, the length scales $\ell_f$ and $\ell_i$ are initilised at random from $\{10,20,30,40,80\}$, $\alpha_i$ and $\sigma_i$ are drawn from $\mathcal{U}[0,1]$ with the value of $\sigma_i$ multiplied by 100 to avoid local minima that would under-estimate the variance. The delays are inilialised at random with the more realistic speed bounds $s_{min}$=500 bp per min and $s_{max}$ = 5kb per min when an ensemble of cells is considered. The parameters are then freely optimised with the bounds given in Table \ref{tab:bounds}.

\subsubsection*{Parameter gradients}
To obtain ML estimates of the parameters we maximise the log marginal likelihood. To do this we require the gradients of the covariance function w.r.t the parameters. The gradients w.r.t $\alpha_i$ and $\sigma_i$ are straight forward. Here we give the expressions for the gradients of $\mathsf{cov}[y_i(t),y_j(t')] = K_{yy}$ w.r.t $\ell_f$, $\ell_i$ and $D_i$. We have

\begin{eqnarray}\nonumber
\frac{\partial K_{yy} }{\partial \ell_f}&=& \alpha_i\alpha_j\frac{\sigma_f^2(\ell_i^2+\ell_j^2)}{(\ell_f^2+\ell_i^2+\ell_j^2)^{\frac{3}{2}}}\exp\Bigg(-\frac{(t'-t+D_i-D_j)^2}{2(\ell_f^2+\ell_i^2+\ell_j^2)}\Bigg)\\
&+& \alpha_i\alpha_j\frac{\sigma_f^2\ell_f}{\sqrt{\ell_f^2+\ell_i^2+\ell_j^2}}\exp\Bigg(-\frac{(t'-t+D_i-D_j)^2}{2(\ell_f^2+\ell_i^2+\ell_j^2)}\Bigg)\frac{(t'-t+D_i-D_j)^2}{(\ell_f^2+\ell_i^2+\ell_j^2)^2}
\end{eqnarray}

\begin{eqnarray}\nonumber
\frac{\partial K_{yy} }{\partial \ell_i}&=& -\alpha_i\alpha_j\frac{\sigma_f^2\ell_f\ell_i}{(\ell_f^2+\ell_i^2+\ell_j^2)^{\frac{3}{2}}}\exp\Bigg(-\frac{(t'-t+D_i-D_j)^2}{2(\ell_f^2+\ell_i^2+\ell_j^2)}\Bigg)\\
&+& \alpha_i\alpha_j\frac{\sigma_f^2\ell_f}{\sqrt{\ell_f^2+\ell_i^2+\ell_j^2}}\exp\Bigg(-\frac{(t'-t+D_i-D_j)^2}{2(\ell_f^2+\ell_i^2+\ell_j^2)}\Bigg)\frac{\ell_i(t'-t+D_i-D_j)^2}{(\ell_f^2+\ell_i^2+\ell_j^2)^2}
\end{eqnarray}

\begin{equation}
\frac{\partial K_{yy} }{\partial D_i}=-\alpha_i\alpha_j\frac{\sigma_f^2\ell_f}{\sqrt{\ell_f^2+\ell_i^2+\ell_j^2}}\exp\Bigg(-\frac{(t'-t+D_i-D_j)^2}{2(\ell_f^2+\ell_i^2+\ell_j^2)}\Bigg)\frac{(t'-t+D_i-D_j)}{(\ell_f^2+\ell_i^2+\ell_j^2)}
\end{equation}

To obtain gradient w.r.t the transformed parameters given by equation \ref{eqn:sig}, we employ the chain rule.

\begin{eqnarray}\nonumber
\frac{\partial K_{yy} }{\partial \gamma}&=&\frac{\partial K_{yy} }{\partial \theta}\frac{\partial \theta }{\partial \gamma}\\
&=&\frac{\partial K_{yy} }{\partial \theta}\frac{\exp(\gamma)(\theta_{max}-\theta_{min})}{(1+\exp(\gamma))^2}
\end{eqnarray}

\subsection*{Canonical Pathway and Gene Ontology Analysis}
To determine the biological significance of the 383 genes found to fit the pol-II dynamics model well, we used the Genomatix Pathway System (GePS) to look for enriched canonical pathways and gene ontology categories. Table \ref{tab:can_path} shows the significant canonical pathways ($p$-value $<0.05$) and the observed genes. It is interesting to note that the pair of genes \textit{JAK1} and \textit{JAK2} are responsible for a large number of the significant canonical pathways. These genes have previously been suggested as potential drug targets in breast cancer (see for example \cite{DrugTargets_2012}). The enrichment of the FOXA1 transcriptional network provides further confirmation that our model identifies biologically relevant genes. In recent work, Hurtado \textit{et al.} \cite{Hurtado_2011} showed that FOXA1 influences the interaction of ER$\alpha$ and chromatin and therefore influences the response of breast cancer cells to E2.
Genes in the FOXA1 canonical network found to fit the pol-II model well include \textit{NRIP1} which is believed to be a direct E2 target that mediates the repression of ER$\alpha$ target genes later in the time course\cite{Carroll_2006,jagannathan2011met}.
 Table \ref{tab:go_an} shows the top 20 significant gene ontology terms ($p$-value $<0.05$) for molecular function.

\begin{center}
\begin{table}[ht!]
\begin{center}
\begin{tabular}{|c|c|} 
\hline 
Canonical pathway&Genes\\
\hline
IL-6 signaling pathway(JAK1 JAK2 STAT3) & 	JAK1, JAK2\\
\hline
IFN gamma signaling pathway &	JAK1, JAK2\\
\hline
Proteasome complex	&PSME1, PSMA4, PSMB5, PSMA2\\
\hline
IL-3 signaling pathway(JAK1 JAK2 STAT5)& 	JAK1, JAK2\\
\hline
Stat3 signaling pathway	&JAK1, JAK2\\
\hline
FOXA1 transcription factor network&	AP1B1, NDUFV3, NRIP1, SHH\\
\hline
PDGFR-alpha signaling pathway&	JAK1, PDGFB, SHB\\
\hline
Hypoxia and p53 in the cardiovascular system&	FHL2, HIF1A, GADD45A\\
\hline
LIF signaling pathway &	JAK1, JAK2\\
\hline
IL-5 signaling pathway&	JAK1, JAK2\\
\hline
p53 signaling pathway&	TIMP3, GADD45A\\
\hline
IL-10 anti-inflammatory signaling pathway&	JAK1, BLVRB\\
\hline
AndrogenReceptor&	SPDEF, FHL2, STUB1\\
& NCOR2, NRIP1\\
\hline
Integrin signaling pathway&	CSK, ACTN1, NOLC1\\
\hline
Erythropoietin mediated &\\
 neuroprotection through NF-KB&	HIF1A, JAK2\\
\hline
PDGFR-beta signaling pathway&	ACTR2, HCK, CSK, \\
&PDGFB, CTTN, JAK2\\
\hline
Mechanisms of transcriptional &\\
repression by dna methylation&	RBBP7, MBD1\\
\hline
Hypoxia-inducible factor in &\\
the cardivascular system&	HIF1A, LDHA\\
\hline
\end{tabular} 
\end{center}
\caption{Significant canonical pathways ($p$-value $<0.05$) for the 383 genes found to fit the pol-II dynamics model well.}
\label{tab:can_path}
\end{table} 
\end{center}

\begin{center}
\begin{table}[ht!]
\begin{center}
\begin{tabular}{|c|} 
\hline 
Molecular function \\ 
\hline
Structural constituent of ribosome\\
RNA binding\\
Methyl-CpG binding\\
Protein binding\\
Structural molecule activity\\
Nucleic acid binding\\
rRNA binding\\
Non-membrane spanning protein tyrosine kinase activity\\
Ribosomal small subunit binding\\
Pseudouridine synthase activity\\
S100 alpha binding\\
Growth hormone receptor binding\\
Isomerase activity\\
Glucocorticoid receptor binding\\
Translation factor activity, nucleic acid binding\\
NF-kappaB binding\\
Threonine-type peptidase activity\\
Threonine-type endopeptidase activity\\
Intramolecular transferase activity\\
\hline
\end{tabular} 
\end{center}
\caption{Top 20 significant gene ontology terms  ($p$-value $<0.05$) for the 383 genes found to fit the pol-II dynamics model well.
}
\label{tab:go_an}
\end{table} 
\end{center}

Table \ref{tab:pathway_gw} shows the significant canonical pathways ($p$-value $<0.01$) and the observed genes in each of the 12 promoter profile clusters. 
We also perform a gene ontology analysis of the 12 promoter profile clusters using the DAVID tool from the NIH \cite{DAVID_Huang_08,DAVID_Huang_09}. The enriched gene ontology categories ($p$-value $<0.05$) are shown in Table \ref{tab:DAVID_GO_1}, (for molecular function), Table \ref{tab:DAVID_GO_2} (for biological processes) and Table \ref{tab:DAVID_GO_3} (for cellular components). 

\begin{center}
\begin{table}[ht!]
\begin{center}
\begin{tabular}{|c|c|c|} 
\hline 
Cluster&Canonical pathway&Genes\\ 
\hline
1 (37)&PDGFR-beta signaling pathway&	PDGFB, ACTR2, HCK\\
\hline
2 (47)&	-&-\\
\hline
3 (18)&-&-\\
\hline
4 (29)&	Nuclear receptors coordinate the activities&NCOR2, TAF5\\
 &of chromatin remodeling complexes and coactivators&\\
& to facilitate initiation of transcription in carcinoma cells&	\\
\hline
5 (27)&-&-\\
\hline
6 (40)&	-&-\\
\hline
7 (24)&Proteasome complex&	PSMB5, PSME1\\
&Antigen processing and presentation &	PSMB5\\
\hline
8 (47)&- &-\\
\hline
9 (26)&	-&-\\
\hline
10 (38)&IFN gamma signaling pathway&	JAK2, JAK1\\
&IL-6 signaling pathway& 	JAK2, JAK1\\
&IL-3 signaling pathway& 	JAK2, JAK1\\
&Stat3 signaling pathway&	JAK2, JAK1\\
&LIF signaling pathway &	JAK2, JAK1\\
&IL-5 signaling pathway&	JAK2, JAK1\\
&PDGFR-alpha signaling pathway&	SHB, JAK1\\
&IL27-mediated signaling events&	JAK2, JAK1\\
&Role of ErbB2 in signal transduction and oncology&	JAK2, JAK1\\
&IL6-mediated signaling events&	JAK2, JAK1\\
&JAK\_STAT\_MolecularVariation\_2&	JAK2, JAK1\\
\hline
11 (13)&-&-\\
\hline
12 (37)&-&-\\
\hline
\end{tabular} 
\end{center}
\caption{Pathway analysis of clusters from inferred promoter activity profiles. The number in parentheses in column 1 is the cluster size.
}
\label{tab:pathway_gw}
\end{table} 
\end{center}

\begin{center}
\begin{table}[ht!]
\begin{center}
\begin{tabular}{|c|c|l|} 
\hline 
Cluster&GO ID&GO TERM\\ 
\hline
1 (37)&GO:0008092&Cytoskeletal protein binding\\
&GO:0003779&Actin binding\\
&GO:0005085&Guanyl-nucleotide exchange factor activity\\
\hline
2 (47)&	GO:0003723&RNA binding\\
\hline
3 (18)&-&-\\
\hline
4 (29)&	GO:0003723&RNA binding\\
&GO:0030528&transcription regulator activity\\
&GO:0003677&DNA binding\\
&GO:0003700&Transcription factor activity\\
\hline
5 (27)&-&-\\
\hline
6 (40)&	GO:0003735&Structural constituent of ribosome\\
\hline
7 (24)&GO:0003735&Structural constituent of ribosome\\
&GO:0005198&Structural molecule activity\\
&GO:0003723&RNA binding\\
\hline
8 (47)&- &-\\
\hline
9 (26)&GO:0043021&Ribonucleoprotein binding\\
\hline
10 (38)&GO:0005131&Growth hormone receptor binding\\
&GO:0051427&Hormone receptor binding\\
&GO:0032553&Ribonucleotide binding\\
&GO:0032555&Purine ribonucleotide binding\\
&GO:0017076&Purine nucleotide binding\\
&GO:0005525&GTP binding\\
&GO:0019001&Guanyl nucleotide binding\\
&GO:0032561&Guanyl ribonucleotide binding\\
&GO:0004713&Protein tyrosine kinase activity\\
\hline
11 (13)&-&-\\
\hline
12 (37)&GO:0003735&Structural constituent of ribosome\\
&GO:0005198&Structural molecule activity\\
&GO:0003723&RNA binding\\
\hline
\end{tabular} 
\end{center}
\caption{Enriched gene ontology categories for molecular function ($p$-value $<0.05$) of clusters from inferred promoter activity profiles. The number in parentheses in column 1 is the cluster size.
}
\label{tab:DAVID_GO_1}
\end{table} 
\end{center}

\clearpage
\begin{center}
\begin{longtable}{|c|c|l|}
\hline 
Cluster&GO ID&GO TERM\\ 
\hline
1 (37)&GO:0030036&Actin cytoskeleton organization\\
&GO:0030029&Actin filament-based process\\
&GO:0007010&Cytoskeleton organization\\
&GO:0007517&Muscle organ development\\
&GO:0001503&Ossification\\
&GO:0001501&Skeletal system development\\
&GO:0060348&Bone development\\
&GO:0060537&Muscle tissue development\\
&GO:0051496&Positive regulation of stress fiber formation\\
&GO:0007167&Enzyme linked receptor protein signaling pathway\\
&GO:0045935&Positive regulation of nucleobase,\\
&& nucleoside, nucleotide and nucleic acid metabolic process\\
&GO:0032233&Positive regulation of actin filament bundle formation\\
&GO:0051173&Positive regulation of nitrogen compound metabolic process\\
&GO:0010557&Positive regulation of macromolecule biosynthetic process\\
&GO:0031328&Positive regulation of cellular biosynthetic process\\
&GO:0009891&Positive regulation of biosynthetic process\\
&GO:0051492&Regulation of stress fiber formation\\
&GO:0048008&Platelet-derived growth factor receptor signaling pathway\\
&GO:0032231&Regulation of actin filament bundle formation\\
&GO:0055010&Ventricular cardiac muscle morphogenesis\\
&GO:0055008&Cardiac muscle tissue morphogenesis\\
&GO:0060415&Muscle tissue morphogenesis\\
\hline
2 (47)&	GO:0051272&Positive regulation of cell motion\\
&GO:0043085&Positive regulation of catalytic activity\\
&GO:0044093&Positive regulation of molecular function\\
\hline
3 (18)&GO:0006364&rRNA processing\\
&GO:0016072&rRNA metabolic process\\
\hline
4 (29)&	GO:0010558&Negative regulation of macromolecule biosynthetic process\\
&GO:0031327&Negative regulation of cellular biosynthetic process\\
&GO:0006350&Transcription\\
&GO:0009890&Negative regulation of biosynthetic process\\
&GO:0010605&Negative regulation of macromolecule metabolic process\\
&GO:0016481&Negative regulation of transcription\\
&GO:0010629&Negative regulation of gene expression\\
&GO:0045934&Negative regulation of nucleobase, \\
&&nucleoside, nucleotide and nucleic acid metabolic process\\
&GO:0051172&Negative regulation of nitrogen compound metabolic process\\
\hline
5 (27)&-&-\\
\hline
6 (40)&	GO:0048147&Negative regulation of fibroblast proliferation\\
&GO:0022613&Ribonucleoprotein complex biogenesis\\
\hline
7 (24)&GO:0019941&Modification-dependent protein catabolic process\\
&GO:0043632&Modification-dependent macromolecule catabolic process\\
&GO:0051603&Proteolysis involved in cellular protein catabolic process\\
&GO:0044257&Cellular protein catabolic process\\
&GO:0030163&Protein catabolic process\\
&GO:0006412&Translation\\
&GO:0043161&Proteasomal ubiquitin-dependent protein catabolic process\\
&GO:0010498&Proteasomal protein catabolic process\\
&GO:0044265&Cellular macromolecule catabolic process\\
&GO:0009057&Macromolecule catabolic process\\
&GO:0006508&Proteolysis\\
&GO:0006511&Ubiquitin-dependent protein catabolic process\\
\hline
8 (47)&GO:0042273&Ribosomal large subunit biogenesis\\
&GO:0006396&RNA processing\\
&GO:0006400&tRNA modification\\
\hline
9 (26)&GO:0043086&Negative regulation of catalytic activity\\
\hline
10 (38)&GO:0007242&Intracellular signaling cascade\\
&GO:0015031&Protein transport\\
&GO:0045184&Establishment of protein localization\\
&GO:0008104&Protein localization\\
&GO:0001525&Angiogenesis\\
&GO:0010876&Lipid localization\\
\hline
11 (13)&-&-\\
\hline
12 (37)&GO:0006412&Translation\\
&GO:0006414&Translational elongation\\
&GO:0051168&Nuclear export\\
&GO:0042274&Ribosomal small subunit biogenesis\\
&GO:0000278&Mitotic cell cycle\\
&GO:0006974&Response to DNA damage stimulus\\
&GO:0006913&Nucleocytoplasmic transport\\
&GO:0051169&Nuclear transport\\
&GO:0022613&Ribonucleoprotein complex biogenesis\\
\hline
\caption{Enriched gene ontology categories for biological processes ($p$-value $<0.05$) of clusters from inferred promoter activity profiles. The number in parentheses in column 1 is the cluster size.
}
\label{tab:DAVID_GO_2}
\end{longtable} 
\end{center}

\clearpage
\begin{center}
\begin{longtable}{|c|c|l|}
\hline 
Cluster&GO ID&GO TERM\\ 
\hline
1 (37)&GO:0015629&Actin cytoskeleton\\
&GO:0005856&Cytoskeleton\\
&GO:0043228&Non-membrane-bounded organelle\\
&GO:0043232&Intracellular non-membrane-bounded organelle\\
&GO:0030017&Sarcomere\\
&GO:0030016&Myofibril\\
&GO:0044449&Contractile fiber part\\
&GO:0043292&Contractile fiber\\
&GO:0001725&Stress fiber\\
\hline
2 (47)&	-&-\\
\hline
3 (18)&-&-\\
\hline
4 (29)&	GO:0016604&Nuclear body\\
&GO:0005654&Nucleoplasm\\
&GO:0030529&Ribonucleoprotein complex\\
&GO:0044451&Nucleoplasm part\\
&GO:0031981&Nuclear lumen\\
&GO:0022625&Cytosolic large ribosomal subunit\\
\hline
5 (27)&GO:0030529&Ribonucleoprotein complex\\
&GO:0005732&Small nucleolar ribonucleoprotein complex\\
&GO:0043232&Intracellular non-membrane-bounded organelle\\
&GO:0043228&Non-membrane-bounded organelle\\
\hline
6 (40)&	GO:0044429&Mitochondrial part\\
&GO:0070013&Intracellular organelle lumen\\
&GO:0043233&Organelle lumen\\
&GO:0031974&Membrane-enclosed lumen\\
&GO:0005743&Mitochondrial inner membrane\\
&GO:0019866&Organelle inner membrane\\
&GO:0044455&Mitochondrial membrane part\\
&GO:0033279&Ribosomal subunit\\
&GO:0031966&Mitochondrial membrane\\
&GO:0005739&Mitochondrion\\
&GO:0005740&Mitochondrial envelope\\
&GO:0005840&Ribosome\\
\hline
7 (24)&GO:0005840&Ribosome\\
&GO:0033279&Ribosomal subunit\\
&GO:0030529&Ribonucleoprotein complex\\
&GO:0000313&Organellar ribosome\\
&GO:0005761&Mitochondrial ribosome\\
\hline
8 (47)&GO:0031981&Nuclear lumen\\
&GO:0005730&Nucleolus\\
&GO:0070013&Intracellular organelle lumen\\
&GO:0043233&Organelle lumen\\
&GO:0031974&Membrane-enclosed lumen\\
&GO:0030529&Ribonucleoprotein complex\\
\hline
9 (26)&GO:0031981&Nuclear lumen\\
\hline
10 (38)&GO:0009898&Internal side of plasma membrane\\
&GO:0044459&Plasma membrane part\\
\hline
11 (13)&GO:0022625&Cytosolic large ribosomal subunit\\
&GO:0015934&Large ribosomal subunit\\
&GO:0022626&Cytosolic ribosome\\
\hline
12 (37)&GO:0005840&Ribosome\\
&GO:0033279&Ribosomal subunit\\
&GO:0030529&Ribonucleoprotein complex\\
&GO:0043232&Intracellular non-membrane-bounded organelle\\
&GO:0043228&Non-membrane-bounded organelle\\
&GO:0044445&Cytosolic part\\
&GO:0005761&Mitochondrial ribosome\\
&GO:0000313&Organellar ribosome\\
&GO:0015935&Small ribosomal subunit\\
&GO:0015934&Large ribosomal subunit\\
&GO:0031980&Mitochondrial lumen\\
&GO:0005759&Mitochondrial matrix\\
&GO:0022626&Cytosolic ribosome\\
&GO:0005829&Cytosol\\
&GO:0070013&Intracellular organelle lumen\\
&GO:0043233&Organelle lumen\\
&GO:0031974&Membrane-enclosed lumen\\
&GO:0005739&Mitochondrion\\
&GO:0000315&Organellar large ribosomal subunit\\
&GO:0005762&Mitochondrial large ribosomal subunit\\
\hline
\caption{Enriched gene ontology categories for cellular components ($p$-value $<0.05$) of clusters from inferred promoter activity profiles. The number in parentheses in column 1 is the cluster size.
}
\label{tab:DAVID_GO_3}
\end{longtable} 
\end{center}

\subsection*{Clustering the raw ChIP-Seq reads}
\begin{figure}[ht!]
\centering
\includegraphics[width=0.75\textwidth]{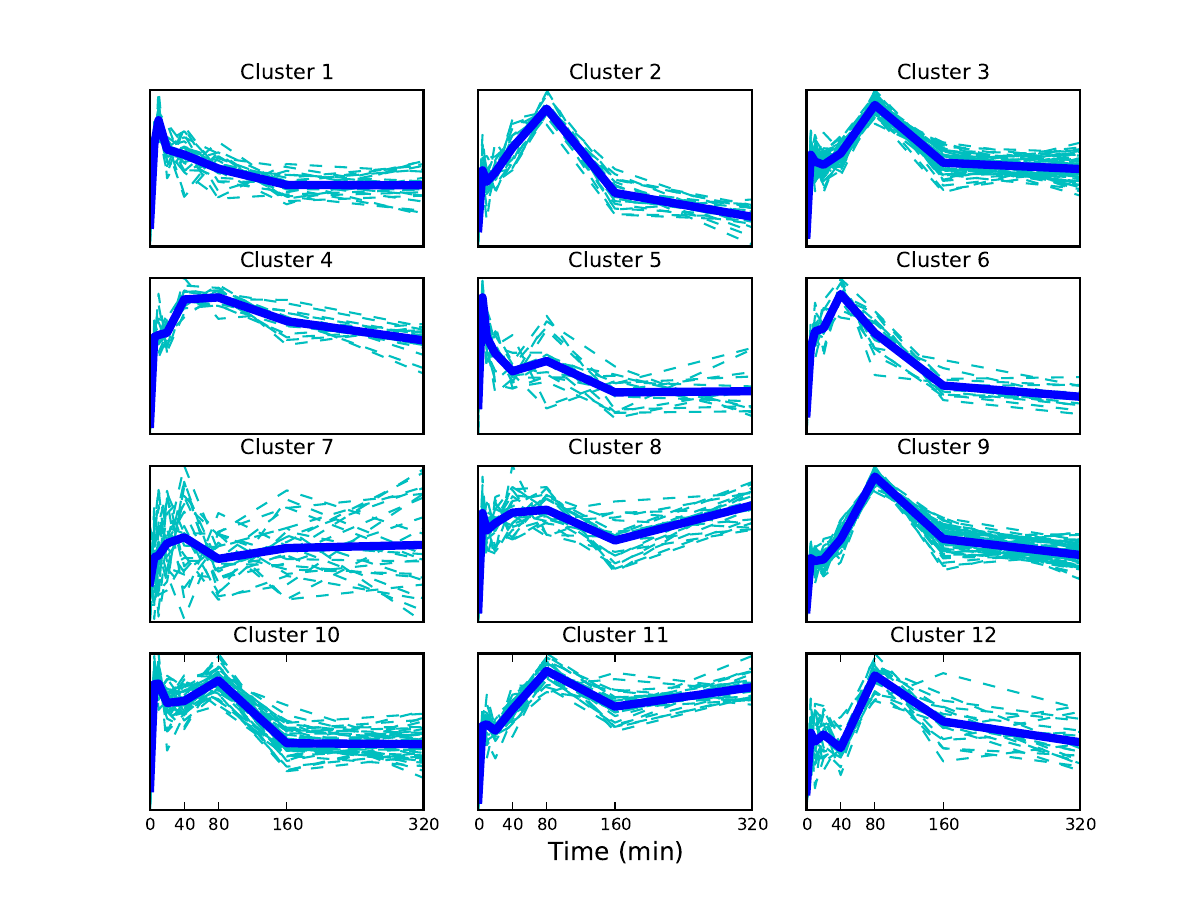}
\caption{Clusters of promoter activity profiles derived directly from the raw ChIP-seq reads. The mean profile in each cluster is shown by the bold line.}
\label{fig:clusters_pp}
\end{figure}

Pol-II occupancy in the proximal promoter region -250 bp to +750 bp relative to the transcription start site (TSS) was computed in RPM for the 383 genes and the time series grouped into 12 clusters. The clusters are shown in Figure \ref{fig:clusters_pp}. Table \ref{tab:pathway_pp} shows the significant canonical pathways ($p$-value $<0.01$) and the observed genes in each of the 12 clusters.  We find that in this case \textit{JAK1} and \textit{JAK2} appear in different clusters which have diffent temporal profiles. This may be due to the noisy nature of the data and the inclusion of paused pol-II in the proximal region time series. Our model which has the potential to uncover the signal due to pol-II that is engaged in transcription could be useful in uncovering relationships which may be missed if we only consider the raw ChIP-seq reads.

\begin{center}
\begin{table}[ht!]
\begin{center}
\begin{tabular}{|c|c|c|} 
\hline 
Cluster&Canonical pathway&Genes\\ 
\hline
1 (24)&Transcriptional activation of dbpB from mRNA&	PDGFB\\
\hline
2 (23)&	-&-\\
\hline
3 (75)&Hypoxia and p53 in the cardiovascular system&	GADD45A, HIF1A\\
\hline
4 (18)&	Generation of amyloid b-peptide by ps1&	ATP5G3\\
\hline
5 (16)&PDGFR-alpha signaling pathway&	SHB, JAK1\\
&IFN gamma signaling pathway&	JAK1\\
&IL-6 signaling pathway&JAK1\\
&IL-10 signaling pathway&JAK1\\
\hline
6 (15)&	-&-\\
\hline
7 (24)&-&-\\
\hline
8 (24)&- &-\\
\hline
9 (67)&Proteasome complex	&PSME1, PSMB5, PSMA4\\
\hline
10 (49)&TPO signaling pathway&	JAK2\\
\hline
11 (29)&Glypican 3 network&	SHH\\
&Sonic hedgehog receptor ptc1 regulates cell cycle	&SHH\\
\hline
12 (19)&Hypoxia-inducible factor in the cardivascular system&	LDHA\\
&Fibrinolysis pathway&	ATP2A2\\
\hline
\end{tabular} 
\end{center}
\caption{Pathway analysis of clusters from raw ChIP-seq reads in the proximal promoter region -250bp to +750bp from the TSS. The number in parentheses in column 1 is the cluster size.
}
\label{tab:pathway_pp}
\end{table} 
\end{center}

\subsection*{Transcription Factor Binding}
\subsubsection*{Motifs}

Tullai \textit{et al.} \cite{Tullai_07} investigated genes that are co-regulated by shared transcription factor binding sites (TFBS). In particular, they found certain TFBS were enriched in the promoters of early response genes. We therefore investigated whether the promoters of genes in the different promoter profile clusters are enriched for different TFs. We use Pscan \cite{ZambelliPP09} to look for enriched TF motifs among those available in JASPAR \cite{Jaspar_04}.
The proximal promoter region
 -450 bp to +50 bp relative to the TSS of the genes in each cluster was analyzed. Table \ref{tab:tfbs} shows signifiantly enriched TFBS in each cluster ($p$-value $<0.05$). Shown are TFs whose binding
sites are over-represented in at least 5 clusters.
 The estrogen response element (ERE)
is enriched in five clusters (1, 2, 5, 6 and 10), indicating that our modelling identifies estrogen 
responsive regions. The clusters containing an ERE have mean promoter
activity profiles with distinct early peaks followed by decrease in activity which suggests
transient activity. Additionally, clusters 1, 2 and 10 have relatively early peaks.

\begin{center}
\begin{table}[ht!]
\begin{center}
\begin{tabular}{|c|c|c|c|c|c|c|c|c|c|c|c|c|}
\hline 
TF&\multicolumn{12}{|c|}{Cluster}\\ \cline{2-13} 
&1&2&3&4&5&6&7&8&9&10&11&12\\
\hline 
GABPA&$\checkmark$&$\checkmark$&$\checkmark$&$\checkmark$&$\checkmark$&-&$\checkmark$&$\checkmark$&$\checkmark$&-&$\checkmark$&$\checkmark$\\
\hline
Zfx&$\checkmark$&$\checkmark$&-&$\checkmark$&$\checkmark$&-&-&$\checkmark$&$\checkmark$&$\checkmark$&$\checkmark$&$\checkmark$\\
\hline
Klf4&$\checkmark$&$\checkmark$&-&$\checkmark$&-&-&$\checkmark$&$\checkmark$&$\checkmark$&$\checkmark$&-&$\checkmark$\\
\hline
ELK1&$\checkmark$&-&-&$\checkmark$&$\checkmark$&-&$\checkmark$&$\checkmark$&$\checkmark$&-&$\checkmark$&$\checkmark$\\
\hline
HIF1A::ARNT&$\checkmark$&$\checkmark$&-&$\checkmark$&$\checkmark$&$\checkmark$&$\checkmark$&-&-&$\checkmark$&$\checkmark$&-\\
\hline
ELK4&$\checkmark$&-&$\checkmark$&$\checkmark$&$\checkmark$&-&$\checkmark$&$\checkmark$&$\checkmark$&-&-&$\checkmark$\\
\hline
SP1&$\checkmark$&$\checkmark$&-&$\checkmark$&-&-&-&$\checkmark$&$\checkmark$&$\checkmark$&-&$\checkmark$\\
\hline
TFAP2A&$\checkmark$&$\checkmark$&-&$\checkmark$&-&$\checkmark$&-&$\checkmark$&-&$\checkmark$&-&-\\
\hline
Mycn&$\checkmark$&-&-&$\checkmark$&$\checkmark$&-&-&$\checkmark$&-&$\checkmark$&$\checkmark$&-\\
\hline
Myc&$\checkmark$&-&-&$\checkmark$&$\checkmark$&-&-&$\checkmark$&-&$\checkmark$&$\checkmark$&-\\
\hline
Pax5&$\checkmark$&$\checkmark$&-&$\checkmark$&-&-&-&-&-&$\checkmark$&-&$\checkmark$\\
\hline
\textbf{ER$\alpha$}&$\checkmark$&$\checkmark$&-&-&$\checkmark$&$\checkmark$&-&-&-&$\checkmark$&-&-\\
\hline
Arnt::Ahr&-&$\checkmark$&-&$\checkmark$&-&-&$\checkmark$&$\checkmark$&-&$\checkmark$&-&-\\
\hline
\end{tabular} 
\end{center}
\caption{Significantly over-represented ($p$-value $<0.05$) transcription factor binding sites in the promoter profile clusters. We use Pscan to look for enriched TF motifs among those available in JASPAR.
The proximal promoter region
 -450 bp to +50 bp relative to the TSS of the genes in each cluster was analyzed.}
\label{tab:tfbs}
\end{table} 
\end{center}

\begin{figure}[ht]
\centering
\includegraphics[width=0.75\textwidth]{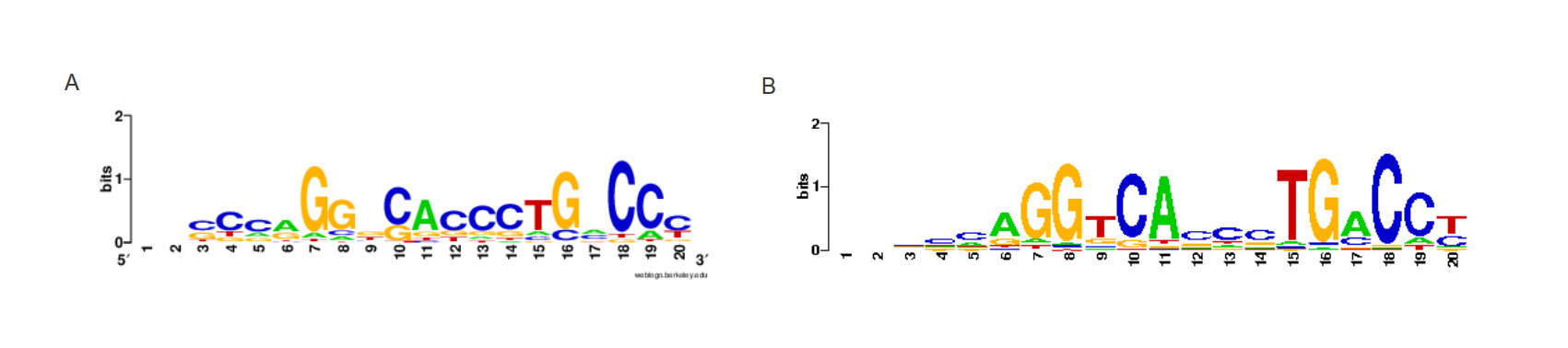}
\caption{Consensus sequence of regions matching the ERE motif in the promoter profile clusters enriched for the ERE motif (A) and the Estrogen Response Element (B).}
\label{fig:motifs}
\end{figure}

Next we investigated the EREs in the genes belonging to the 5 clusters enriched for the ERE motif. For each promoter sequence, the best sequence match to the ERE position frequency matrix (PFM) in JASPAR (MA0112.2) was determined. We keep those sequences with a matrix score greater than the mean score for matches found in the promoter sequences over the whole genome (For the ERE PFM this value is $0.73$ when we consider the region
 -450 bp to +50 bp relative to the TSS). We used these sequences to determine the consensus ERE motif in this group of genes. To determine the consensus sequence, we report a single nucleotide for a given position if the nucleotide has a frequency greater than 50\% and a frequency twice as large as the next nucleotide. We obtain a consensus sequence of 5'-GGnCACCCTGnCC-3' (where n is any nucleotide) and an average matrix score of 0.77. The sequence is visualised in Figure \ref{fig:motifs} (A). The sequence of the ERE is known and given as  5'-GGTCAnnnTGACC-3' \cite{Klinge_2001,Welboren_Stunnenberg_Sweep_Span_2007}. The sequence corresponding to the PFM MA0112.2 is visualised in Figure \ref{fig:motifs} (B). We see that the ERE motif we obtain agrees well with the known motif.

Table \ref{tab:tf_motifs} shows the EREs in each of the 5 clusters visualised using WebLogo. We also show the consensus sequence and the average matrix score. To determine the consensus sequence, we report a single nucleotide for a given position if the nucleotide has a frequency greater than 50\% and a frequency twice as large as the next nucleotide. We see that there is some diversity in the motifs correspoding to different clusters but the consensus sequences agree with the known motif. Differences appear at at most 3 positions with the consensus sequence for cluster 10 differing at only two positions. We see that for the half site  `TGACC' the `A' does not appear in the consensus sequence in all the clusters.  

\begin{center}
\begin{table}[ht!]
\begin{center}
\begin{tabular}{|c|c|c|c|} 
\hline 
Cluster&ERE Motif &Consensus sequence& Average Matrix\\
&&& Score\\ 
\hline
1 &	\includegraphics[scale=.4]{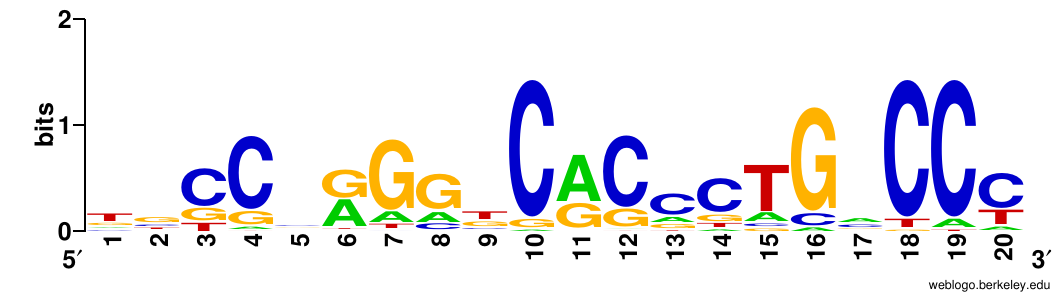}&GnnCACCCTGnCCC&0.772\\
\hline
2 &	\includegraphics[scale=.4]{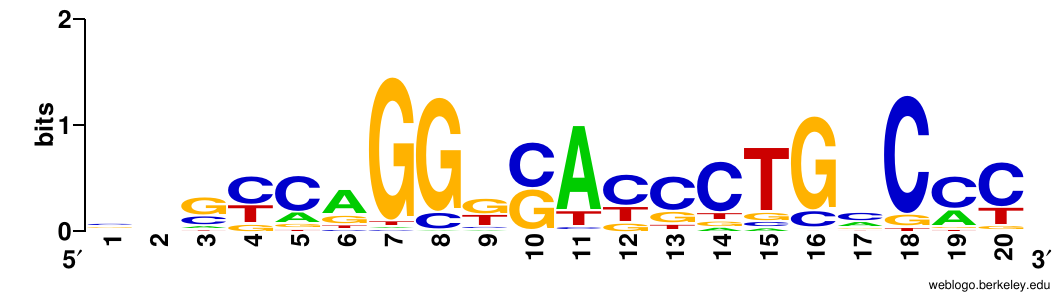}&GGnnACCCTGnCCn&0.77\\
\hline
5 &	\includegraphics[scale=.4]{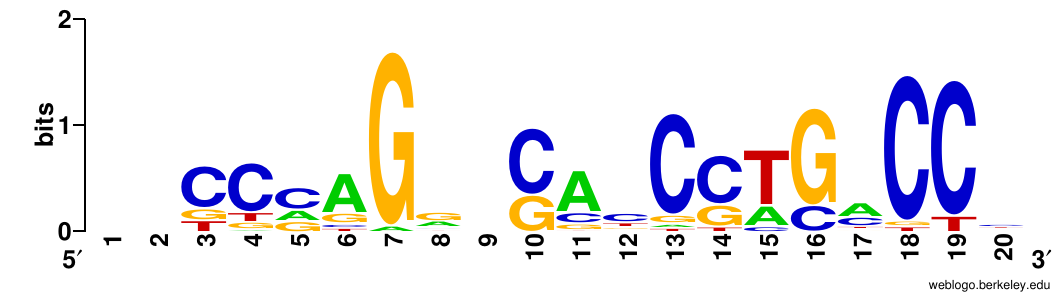}&GGnnAnCCTGnCCn&0.761\\
\hline
6 &	\includegraphics[scale=.4]{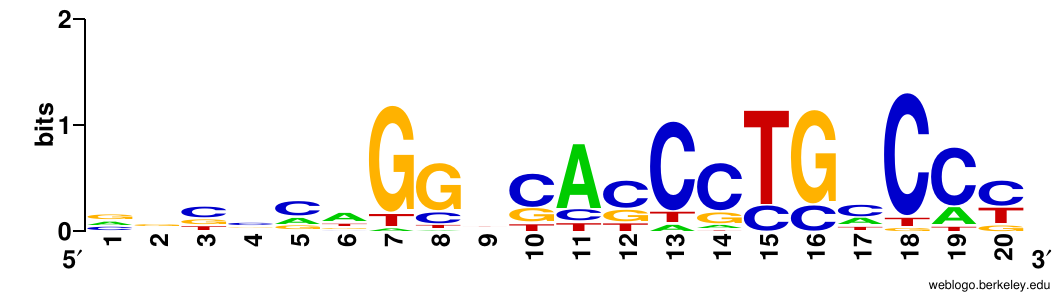}&GGnnACCnTGnCCn&0.762\\
\hline
10 &	\includegraphics[scale=.4]{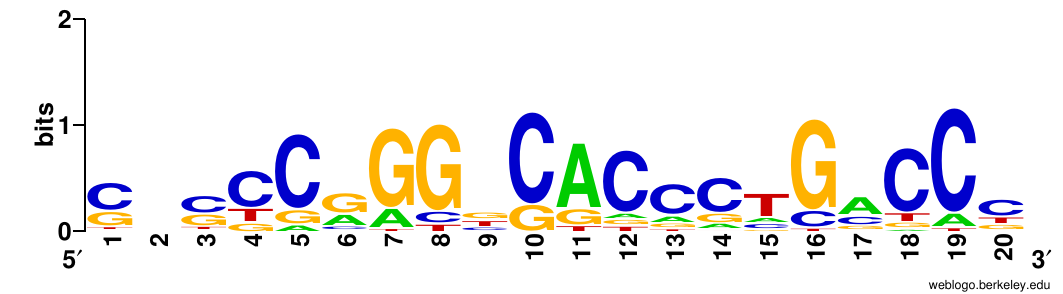}&GGnCACCCTGnCCn&0.765\\
\hline
\end{tabular} 
\end{center}
\caption{Analysis of the ERE in promoter regions of gene clusters obtained from inferred promoter activity profiles. The EREs in each of the 5 clusters are visualised using WebLogo (http://weblogo.berkeley.edu/). The consensus sequence is shown from postion 7 which corresponds to the known ERE motif. The average matrix score is computed using the sequence matrix scores from Pscan.}
\label{tab:tf_motifs}
\end{table} 
\end{center}

\subsubsection*{Transcription factor binding}
Determining the TFBS motifs enriched in each cluster provides a way to determine the influence of TFs on transcription. As a complementary approach, we also investigated the TF peaks in regions ranging from 1 to 100 kb around the gene transcription start site for all genes in each cluster using ChIP-seq data for a number of TFs measured under similar experimental conditions (i.e. MCF-7 breast cancer cells treated with E2) in the cistrome database (\url{http://cistrome.org}).

 Tables \ref{tab:tf_gw1} to \ref{tab:tf_gw5} show the number of genes with TF binding peaks for regions around the TSS ranging from 1 to 100 kb for each cluster for 7 TFs namely ER$\alpha$ \cite{Welboren_2009}, FoxA1 \cite{Lupien2008}, c-Fos \cite{Joseph2010}, c-Jun \cite{Joseph2010}, c-MYC \cite{Hua_2009}, SRC-3 \cite{Lanz01042010}, TRIM24 \cite{Tsai2010}. In the tables, statistically significant ($p$-value $<0.05$) proportions are indicated in red (larger than expected) and green (lower than expected) with associated $p$-values in parentheses. These p-values are obtained empirically by drawing 1e6 samples from a hypergeometric distribution.

We investigated the overlap of the binding sites for ER$\alpha$ and FOXA1 both in the 151 genes belonging to the rapid response genes in clusters 1, 2, 4, and 10 and genome-wide using the peaks obtained from \cite{Welboren_2009} (ER$\alpha$) and \cite{Lupien2008} (FOXA1) and reported in the cistrome database. We investigated regions around the TSS ranging
from 2 to 100 kb. Tables \ref{tab:ER_FOXA1_1}-\ref{tab:ER_FOXA1_4} show the number of ER$\alpha$ and FOXA1 peaks and the overlap. The statistical significance is determined by comparing the overlap in random gene lists of the same size.

\begin{center}
\begin{table}[ht!] 
\begin{center}
\begin{tabular}{|c|c|c|c|c|c|c|c|} 
\hline 
Cluster&\multicolumn{7}{|c|}{TFs}\\ \cline{2-8}
&ER$\alpha$&FOXA1&c-FOS&c-JUN&MYC&SRC-3&TRIM24\\ 
\hline
1 (37)&5& 4& 2& 3& 1& 7& 9 \\
\hline
2 (47)&{\color{red} 9 }(*)& 3& 2& 2& {\color{red} 4 }& {\color{red} 12 }(*)& 10 \\
\hline
3 (18)&3& 2& 2& 1& {\color{red} 3 }(*)& 3& 2 \\
\hline
4 (29)&4& 2& 1& {\color{green} 0 }(***)& {\color{green} 0 }(***)& 3& 5 \\
\hline
5 (27)&3& {\color{green} 0 }(***)& {\color{green} 0 }(***)& {\color{red} 5 }(*)& {\color{red} 4 }(*)& {\color{red} 7 }& 5 \\
\hline
6 (40)&5& 3& 3& {\color{green} 0 }(***)& 3& {\color{red} 8 }& {\color{green} 2 } \\
\hline
7 (24)&1& 2& {\color{green} 0 }(***)& {\color{red} 3 }& 1& {\color{red} 6 }& {\color{red} 7 } \\
\hline
8 (47)&3& 2& 1& 3& {\color{red} 4 }& 6& {\color{red} 14 }(*) \\
\hline
9 (26)&2& 2& {\color{red} 4 }& {\color{red} 5 }(**)& 1& 5& 6 \\
\hline
10 (38)&{\color{red} 9 }(*)& 2& 1& {\color{green} 0 }(***)& {\color{green} 0 }(***)& 3& 9 \\
\hline
11 (13)&{\color{green} 0 }(***)& {\color{green} 0 }(***)& {\color{red} 3 }& 1& 1& 1& 1 \\
\hline
12 (37)&5& {\color{green} 0 }(***)& 2& {\color{red} 5 }(*)& 2& {\color{red} 11 }(**)& 7 \\
\hline
\end{tabular} 
\end{center}
\caption{Analysis of transcription factor binding in 1kbp regions of genes in gene clusters obtained from inferred promoter activity profiles. The number in parentheses in the first column is the cluster size. For each TF, we show the number of genes with peaks. Statistically significant proportions ($p$-value $<0.05$) are indicated in red (larger than expected). For $p$-values less than $0.01$, the associated $p$-values are indicated in parentheses according to the following scale (***: $p<0.0001$,**: $p<0.001$,*:$p<0.01$).}
\label{tab:tf_gw1}
\end{table} 
\end{center}

\begin{center}
\begin{table}[ht!]
\begin{center}
\begin{tabular}{|c|c|c|c|c|c|c|c|} 
\hline 
Cluster&\multicolumn{7}{|c|}{TFs}\\ \cline{2-8}
&ER$\alpha$&FOXA1&c-FOS&c-JUN&MYC&SRC-3&TRIM24\\ 
\hline
1 (37)&{\color{red} 8 }& 4& 3& 3& 1& 8& 10 \\
\hline
2 (47)&{\color{red} 10 }& 3& 3& 2& {\color{red} 5 }(*)& {\color{red} 14 }(**)& 11 \\
\hline
3 (18)&3& 2& 2& 1& {\color{red} 3 }& 3& 3 \\
\hline
4 (29)&4& 2& 1& {\color{green} 0 }(***)& {\color{green} 0 }(***)& 3& {\color{red} 9 } \\
\hline
5 (27)&4& {\color{green} 0 }(***)& 1& {\color{red} 5 }(*)& {\color{red} 6 }(***)& {\color{red} 8 }& 6 \\
\hline
6 (40)&{\color{red} 9 }& {\color{red} 5 }& 5& {\color{green} 0 }(***)& 3& {\color{red} 11 }(*)& {\color{green} 3 } \\
\hline
7 (24)&2& 3& {\color{green} 0 }(***)& {\color{red} 3 }& 1& {\color{red} 7 }& {\color{red} 10 }(*) \\
\hline
8 (47)&5& 2& 1& 4& {\color{red} 4 }& 9& {\color{red} 19 }(**) \\
\hline
9 (26)&3& 2& {\color{red} 6 }(*)& {\color{red} 6 }(**)& 1& {\color{red} 7 }& 7 \\
\hline
10 (38)&{\color{red} 11 }(*)& 3& 2& {\color{green} 0 }(***)& 1& 5& 10 \\
\hline
11 (13)&1& {\color{green} 0 }(***)& {\color{red} 3 }& 1& 1& 1& 3 \\
\hline
12 (37)&6& {\color{green} 0 }(***)& 2& {\color{red} 5 }(*)& 2& {\color{red} 11 }(*)& 8 \\
\hline
\end{tabular} 
\end{center}
\caption{Analysis of transcription factor binding in 2kbp regions.}
\label{tab:tf_gw2}
\end{table} 
\end{center}

\begin{center}
\begin{table}[ht!]
\begin{center}
\begin{tabular}{|c|c|c|c|c|c|c|c|} 
\hline 
Cluster&\multicolumn{7}{|c|}{TFs}\\ \cline{2-8}
&ER$\alpha$&FOXA1&c-FOS&c-JUN&MYC&SRC-3&TRIM24\\ 
\hline
1 (37)&{\color{red} 20 }(*)& 9& 8& 4& 1& {\color{red} 18 }& 22 \\
\hline
2 (47)&{\color{red} 24 }(*)& {\color{red} 13 }& {\color{red} 12 }& 6& {\color{red} 7 }(*)& {\color{red} 30 }(***)& 28 \\
\hline
3 (18)&4& 4& 4& 2& {\color{red} 5 }(*)& 8& 7 \\
\hline
4 (29)&11& 6& 4& 2& 1& 12& 18 \\
\hline
5 (27)&9& 2& 3& {\color{red} 6 }(*)& {\color{red} 8 }(***)& 11& 14 \\
\hline
6 (40)&{\color{red} 22 }(**)& 8& 6& 4& 3& 18& 24 \\
\hline
7 (24)&7& 4& 2& 4& 2& {\color{red} 13 }& {\color{red} 16 } \\
\hline
8 (47)&{\color{red} 21 }& 6& 7& {\color{red} 10 }(*)& {\color{red} 7 }(*)& {\color{red} 28 }(***)& {\color{red} 34 }(**) \\
\hline
9 (26)&10& 4& {\color{red} 8 }& {\color{red} 9 }(***)& 1& 8& {\color{red} 20 }(*) \\
\hline
10 (38)&{\color{red} 26 }(***)& {\color{red} 11 }& 9& {\color{green} 0 }(***)& 1& {\color{red} 21 }(*)& {\color{red} 24 } \\
\hline
11 (13)&4& {\color{green} 0 }(***)& {\color{red} 5 }& 2& 1& 4& 8 \\
\hline
12 (37)&12& {\color{green} 2 }& 7& {\color{red} 10 }(**)& 4& {\color{red} 20 }(*)& 23 \\
\hline
\end{tabular} 
\end{center}
\caption{Analysis of transcription factor binding in 20kbp regions.}
\label{tab:tf_gw3}
\end{table} 
\end{center}

\begin{center}
\begin{table}[ht!]
\begin{center}
\begin{tabular}{|c|c|c|c|c|c|c|c|} 
\hline 
Cluster&\multicolumn{7}{|c|}{TFs}\\ \cline{2-8}
&ER$\alpha$&FOXA1&c-FOS&c-JUN&MYC&SRC-3&TRIM24\\ 
\hline
1 (37)&29& 20& {\color{red} 26 }(***)& {\color{red} 12 }& 4& {\color{red} 32 }(*)& {\color{red} 36 } \\
\hline
2 (47)&{\color{red} 41 }(*)& {\color{red} 26 }& 23& 11& {\color{red} 12 }(*)& {\color{red} 43 }(**)& 43 \\
\hline
3 (18)&{\color{red} 17 }& 7& 10& 6& {\color{red} 6 }& 14& 16 \\
\hline
4 (29)&{\color{red} 29 }(***)& 17& 15& {\color{red} 10 }& 5& {\color{red} 25 }& 28 \\
\hline
5 (27)&21& 8& 11& {\color{red} 12 }(*)& {\color{red} 11 }(**)& 19& 24 \\
\hline
6 (40)&{\color{red} 36 }(*)& 15& 19& 11& 6& {\color{red} 35 }(*)& 38 \\
\hline
7 (24)&15& 11& 8& {\color{red} 9 }& 5& 18& 22 \\
\hline
8 (47)&{\color{red} 42 }(**)& 20& 22& {\color{red} 15 }& 9& {\color{red} 41 }(*)& {\color{red} 45 } \\
\hline
9 (26)&{\color{red} 23 }& 15& {\color{red} 16 }(*)& {\color{red} 12 }(**)& 5& {\color{red} 22 }& 24 \\
\hline
10 (38)&{\color{red} 34 }(*)& {\color{red} 27 }(**)& {\color{red} 20 }& 5& 4& {\color{red} 34 }(*)& 36 \\
\hline
11 (13)&9& 4& 8& 4& 2& 10& 13 \\
\hline
12 (37)&{\color{red} 31 }& {\color{green} 11 }& {\color{red} 19 }& {\color{red} 14 }(*)& 5& 28& 35 \\
\hline
\end{tabular} 
\end{center}
\caption{Analysis of transcription factor binding in 100kbp regions.}
\label{tab:tf_gw5}
\end{table} 
\end{center}

\begin{center}
\begin{table}[ht!]
\begin{center}
\begin{tabular}{|c|c|c|c|} 
\hline 
Genes&\# of ER$\alpha$ peaks& \# of FOXA1 peaks &ER$\alpha$ and FOXA1 overlap\\ 
\hline
Clusters 1, 2, 4, and 10 (151)&28 (12) &11 (6) &7 (0.042)\\
\hline
All genes ($\sim$ 20,000)&1596&758&130 \\
\hline
\end{tabular} 
\end{center}
\caption{Overlap of ER$\alpha$ and FOXA1 binding in a 1 kb region around the TSS. The numbers in parentheses in the first column are the number of genes. In each TF peak column, we show the expected number of peaks in a set of random random genes of the same size in parentheses. In the overlap column the associated p-value is shown in parentheses.}
\label{tab:ER_FOXA1_1}
\end{table} 
\end{center}

\begin{center}
\begin{table}[ht!]
\begin{center}
\begin{tabular}{|c|c|c|c|} 
\hline 
Genes&\# of ER$\alpha$ peaks& \# of FOXA1 peaks &ER$\alpha$ and FOXA1 overlap\\
\hline
Clusters 1, 2, 4, and 10 (151)&36 (17) &13 (7) &8 (0.038)\\
\hline
All genes ($\sim$ 20,000)&2220&929&177 \\
\hline
\end{tabular} 
\end{center}
\caption{Overlap of ER$\alpha$ and FOXA1 binding in a 2 kb region around the TSS.}
\label{tab:ER_FOXA1_2}
\end{table} 
\end{center}

\begin{center}
\begin{table}[ht!]
\begin{center}
\begin{tabular}{|c|c|c|c|} 
\hline 
Genes&\# of ER$\alpha$ peaks& \# of FOXA1 peaks &ER$\alpha$ and FOXA1 overlap\\
\hline
Clusters 1, 2, 4, and 10 (151)&125 (63) &44 (26) &19 (0.045)\\
\hline
All genes ($\sim$ 20,000)&7229&2991&626 \\
\hline
\end{tabular} 
\end{center}
\caption{Overlap of ER$\alpha$ and FOXA1 binding in a 20 kb region around the TSS.}
\label{tab:ER_FOXA1_3}
\end{table} 
\end{center}

\begin{center}
\begin{table}[ht!]
\begin{center}
\begin{tabular}{|c|c|c|c|} 
\hline 
Genes&\# of ER$\alpha$ peaks& \# of FOXA1 peaks &ER$\alpha$ and FOXA1 overlap\\
\hline
Clusters 1, 2, 4, and 10 (151)&488 (254) &171 (100) &66 (0.006)\\
\hline
All genes ($\sim$ 20,000)&17942&7927&1691 \\
\hline
\end{tabular} 
\end{center}
\caption{Overlap of ER$\alpha$ and FOXA1 binding in a 100 kb region around the TSS.}
\label{tab:ER_FOXA1_4}
\end{table} 
\end{center}

\clearpage
\bibliography{bibtex/biology,bibtex/gp,bibtex/delay,../../../bib/lawrence,../../../bib/other}{}

\end{document}